\documentclass[a4paper,fleqn]{cas-dc}
\usepackage[numbers]{natbib}
\usepackage{booktabs}
\usepackage{adjustbox}
\usepackage{longtable}
\usepackage{caption}
\usepackage{mathtools}
\usepackage{amssymb}
\usepackage{amsmath}
\usepackage{etoolbox}

\makeatletter
\pretocmd{\start@align}{\setlength{\mathindent}{0pt}}{}{}
\makeatother

\newlength{\myalignwidth}

\def\tsc#1{\csdef{#1}{\textsc{\lowercase{#1}}\xspace}}
\tsc{WGM}
\tsc{QE}
\tsc{EP}
\tsc{PMS}
\tsc{BEC}
\tsc{DE}
\begin{document}
\let\WriteBookmarks\relax
\def\floatpagepagefraction{1}
\def\textpagefraction{.001}
\let\printorcid\relax

\shorttitle{}

\shortauthors{Chaoqian Wang {\it et~al.}}

\title [mode = title]{The conflict between self-interaction and updating passivity in the evolution of cooperation}                      

\author[1]{Chaoqian Wang}
\ead{CqWang814921147@outlook.com}
\credit{Conceptualization; Methodology; Writing}

\author[2,3]{Wenqiang Zhu}
\credit{Methodology; Validation}

\author[4]{Attila Szolnoki}
\cormark[1]
\cortext[cor1]{Corresponding author}
\ead{szolnoki.attila@ek-cer.hu}
\credit{Conceptualization; Validation; Writing}

\address[1]{Department of Computational and Data Sciences, George Mason University, Fairfax, VA 22030, USA}
\address[2]{School of Mathematical Science, Dalian University of Technology, Dalian 116024, China}
\address[3]{Institute of Artificial Intelligence, Beihang University, Beijing 100191, China}
\address[4]{Institute of Technical Physics and Materials Science, Centre for Energy Research, P.O. Box 49, H-1525 Budapest, Hungary}

\begin{abstract}
In social dilemmas under weak selection, the capacity for a player to exhibit updating passivity or interact with its own strategy can lead to conflicting outcomes. The central question is which effect is stronger and how their simultaneous presence influences the evolution of cooperation. We introduce a model that considers both effects using different weight factors. We derive theoretical solutions for the conditions of cooperation success and the cooperation level under weak selection, scanning the complete parameter space. When the weight factors are equally strong, the promoting effect of self-interaction to cooperation surpasses the inhibitory effect of updating passivity. Intriguingly, however, we identify non-monotonous cooperation-supporting effects when the weight of updating passivity increases more rapidly. Our findings are corroborated by Monte Carlo simulations and demonstrate robustness across various game types, including the prisoner's dilemma, stag-hunt, and snowdrift games.
\end{abstract}

% \includegraphics{figs/grabs.pdf}

\begin{highlights}
\item Examining cooperation levels under weak selection
\item Analyzing conflict between self-interaction and updating passivity
\item Equally strong self-interaction outweighs updating passivity
\item Rapid updating passivity growth reveals non-monotonous cooperation threshold
\item Conclusions consistent across various game types
\end{highlights}


\maketitle

\section{Introduction}\label{secintro}
The study of interactions and microscopic updating dynamics has been crucial for determining the conditions governing the evolutionary outcomes of competing strategies in social dilemmas~\cite{kaiping_pre14,roca_plr09,takesue_pa19}. Over the past two decades, extensive research has been conducted on this topic, leading to the establishment of some generally valid conclusions~\cite{nowak1992evolutionary,perc_pr17}. Specifically, fixed and stable interactions with partners—distinguishing well-mixed from structured populations—enhance direct reciprocity among neighbors~\cite{szabo_pr07,wang_z_epjb15}. Consequently, the term ``network reciprocity'' was proposed to emphasize its vital role in supporting cooperation mechanisms~\cite{nowak_s06}. Intriguingly, variations in an individual's state over time can also yield significant consequences. The vast range of updating rules raises further questions concerning the resilience of cooperation against defection~\cite{szolnoki_pa18,ohtsuki_jtb06,zhu_h_pla20,takesue_epl18,zhu_pc_epjb21}. One might assume that the motivation for an individual to change their state (strategy) depends on the payoff values obtained by their current strategy and the alternatives offered by competitors. However, this hypothesis is not universally applicable, as other factors can also contribute to determining a strategy's fitness~\cite{traulsen_jtb07b,allen2014games}. In such cases, the payoff has a marginal effect on reproductive success, leading to the establishment of the weak selection limit~\cite{tarnita2009evolutionary,maciejewski_pcbi14}. This scenario allows for analytically feasible solutions even in structured populations, making it a popular research direction in recent years~\cite{wild_jtb07,fu_pre09b,debarre2014social,allen2017evolutionary,su2018understanding,su2019spatial,fotouhi2018conjoining,allen2019evolutionary,mcavoy2020social,su2022evolution,su2022evolution_asy}.

In line with this latter assumption, previous research has demonstrated that a certain level of inertia in strategy updates, wherein a player is unwilling to alter their current strategy despite contradicting payoff values, can be detrimental and hinder cooperation~\cite{wang2023inertia}. A similar effect can be achieved by imposing a weight factor dictating the willingness of a strategy change~\cite{wang2023evolution}. In contrast, extending the interaction range of the focal player to include not only nearest neighbors but also their own strategy as an opponent can produce opposite effects~\cite{nowak1992evolutionary,szabo_pre98}. This self-interaction can be particularly justified in biologically inspired ecological systems, where an actor's offspring are in close proximity to the parent. Evidently, this extension benefits cooperators, as cooperator-cooperator interactions yield higher incomes than defector-defector bonds, regardless of the social dilemma's nature. This raises the question of which effect has a more dominant influence on the evolution of cooperation: the negative consequence of updating passivity or the positive effect of self-interaction?

To address this question, we consider a structured population with players distributed on a vertex-transitive graph, where players cannot distinguish their positions by observing the structure of the graph. We introduce two key control parameters that determine the strength of self-interaction and the extent of strategy updating passivity. As technical terms, we may refer to these as the weight of self-gaming and self-learning, respectively, whereby the aforementioned effects can be described as self-loops on interaction and learning graphs~\cite{su2019spatial,debarre2014social,ohtsuki2007breaking}. Our primary objective is to provide analytical results for the critical benefit-to-cost value within the parameter space of these weight factors. In addition to theoretical calculations, which incorporate the identity-by-descent method (IBD)~\cite{allen2014games} and pair approximation~\cite{ohtsuki2006simple}, we also offer numerical simulations to present an overview of system behavior. To assess the robustness of our observations, we investigate all major social dilemmas based on pair interactions of agents, including the prisoner's dilemma, snowdrift, and stag-hunt games~\cite{sigmund_10}.

\section{Model}\label{sec_model}
We capture the essence of a structured population by considering an $L\times L$ square lattice with periodic boundary conditions, hosting $N=L^2$ agents. According to this topology, each agent interacts with $k$ neighbors, which could be von~Neumann ($k=4$) or Moore ($k=8$) neighborhood. As a critical extension, we assume that a player interacts with their own strategy with weight $w_I$, while the interactions with $k$ neighbors are considered with weight $1-w_I$. Similarly, the strategy updating protocol is divided: a player considers their own fitness with weight $w_R$ (referred to as self-learning), while the fitness of neighbors is considered with weight $1-w_R$. 

Based on this hypothesis, we can define the joint transitive interaction and learning graph. The vertex set is denoted by $V$, containing all agents. On the interaction graph, the edge between agent $i$ and $j$ is denoted by $e_{ij}^{[I]}$. According to our assumption, we set $e_{ii}^{[I]}=w_I$, and $e_{ij}^{[I]}=(1-w_I)/k$ if $j$ is one of the $k$ nearest neighbors of player $i$. For other $j$, $e_{ij}^{[I]}=0$. In this way, $\sum_{l\in V}e_{il}^{[I]}=1$ is normalized. The same protocol applies to the learning graph, where the edge between $i$ and $j$ is denoted by $e_{ij}^{[R]}$. Similarly, $e_{ii}^{[R]}=w_R$ marks the self-loop and $e_{ij}^{[R]}=(1-w_R)/k$ connects to the $k$ neighbors. For all other players $j$, $e_{ij}^{[R]}=0$. Moreover, both graphs are indirected, hence $e_{ij}^{[I]}=e_{ji}^{[I]}$ and $e_{ij}^{[R]}=e_{ji}^{[R]}$ for each $ij$ pair. The interaction and learning graphs overlap, with the only difference being the actual values of weight factors $w_I$ and $w_R$ characterizing the self-loops on the graphs.

During an elementary Monte Carlo (MC) step, a random agent $i$ is selected to update the strategy. The strategy of agent $i$ is denoted by $s_i=1$ for cooperation or $s_i=0$ for defection. In the two-player donation game, cooperation means donating $c$ and the other player receiving $b$ where $b>c>0$. A defector player refuses investment but enjoys the benefit of a cooperator partner. Agent $i$'s payoff $\pi_i$ is the average payoff over all games played through the interaction graph,
\begin{equation}\label{eq_pii}
    \pi_i=\sum_{l\in V}e_{il}^{[I]}(-cs_i+bs_l)=-cs_i+b\sum_{l\in V}e_{il}^{[I]}s_l\,.
\end{equation}

After calculating the payoff, we transform it into fitness with form $F_i=\exp{(\delta \pi_i)}$~\cite{mcavoy2020social,wang2023inertia}. The parameter $\delta>0$ represents the strength of selection, and we assume weak selection strength $\delta\to 0^+$ in this work. To define the microscopic dynamics, agent $i$ updates its strategy via the classic death-birth rule through the learning graph. Accordingly, the probability that agent $i$ adopts the strategy of agent $j$ is
\begin{equation}\label{eq_update}
    W(s_i\gets s_j)=\frac{e_{ij}^{[R]}F_j}{\sum_{l\in V}e_{il}^{[R]}F_l}\,.
\end{equation}
This form highlights that the selection process to adopt strategy $s_j$ is proportional to the weighted fitness, where the weight factor is the edge value $e_{ij}^{[R]}$ of the learning graph. To execute a full MC step, the above-described elementary step is repeated $N$ times. In this way, every agent has a chance to update their strategy once on average.

\section{Theoretical analysis}\label{sec_theo}
We assess the system's state by measuring the cooperation level, expressed as the proportion of cooperators in the system. Let the initial cooperation level be $p_C(t_0)=N_C/N$, where $N_C$ denotes the number of cooperators at the initial time $t=t_0$. Cooperation ultimately dominates the system with probability $p_C(t_0)$ under neutral drift (i.e., $\delta=0$, which reduces dynamics to the voter model)~\cite{cox1983occupation,cox1986diffusive}. Therefore, under weak selection ($\delta\to 0^+$), evolution favors cooperation if $\rho_C>p_C(t_0)$, where $\rho_C$ represents the expected final cooperation level over numerous runs at a large $t$. For instance, if the system starts with a single cooperator and $N-1$ defectors, evolution favors cooperation when the final cooperation level $\rho_C>1/N$. Similarly, when the initial cooperation level is $p_C(t_0)\approx 0.5$ in a random state, cooperation is favored when the expected final cooperation level $\rho_C>1/2$.

\subsection{The condition for cooperation success}
For simplicity in analysis, we consider a single initial cooperator, denoted by $1$. Following~\cite{allen2014games,nowak2010evolution}, evolution favors cooperation if the condition in Eq.~(\ref{eq_condi}) is met:
\begin{equation}\label{eq_condi}
    \left\langle\frac{\partial}{\partial\delta}(\mathcal{B}_1-\mathcal{D}_1)\right\rangle_{\begin{smallmatrix}\delta=0\\s_1=1\end{smallmatrix}}>0\,.
\end{equation}
Here, $\langle\cdot\rangle_{\begin{smallmatrix}\delta=0\\s_1=1\end{smallmatrix}}$ signifies the expectation under neutral drift when agent $1$ cooperates. The probabilities of agent $1$ reproducing or replacing its strategy are denoted by $\mathcal{B}_1$ and $\mathcal{D}_1$, respectively.

Considering the standard death-birth updating process described in Eq.~(\ref{eq_update}), agent $1$ reproduces its strategy to another agent $i$ with probability $\mathcal{B}_1$ when agent $i$ is the focal agent and learns agent $1$'s strategy through $W(s_i\gets s_1)$. Conversely, agent $1$'s strategy is replaced with probability $\mathcal{D}_1$ when agent $1$ is the focal agent and learns the strategy of another agent $j$ through $W(s_1\gets s_j)$. Thus, $\mathcal{B}_1$ and $\mathcal{D}_1$ are defined as follows:
\begin{subequations}\label{eq_bd}
	\begin{align}
		\mathcal{B}_1&=\sum_{i\in V}\frac{1}{N} W(s_i\gets s_1) \nonumber\\
		&=\sum_{i\in V}\frac{1}{N}
		\frac{e_{i1}^{[R]}\exp(\delta\pi_1)}{\sum_{l\in V}e_{il}^{[R]}\exp(\delta\pi_l)}\,,
	    \label{eq_b}
		\\
		\mathcal{D}_1&=\frac{1}{N}\sum_{j\in V}W(s_1\gets s_j) \nonumber\\
		&=\frac{1}{N}\sum_{j\in V}\frac{e_{1j}^{[R]}\exp(\delta\pi_j)}{\sum_{l\in V}e_{1l}^{[R]}\exp(\delta\pi_l)}\,.
	    \label{eq_d}
	\end{align}
\end{subequations}
By substituting Eqs.~(\ref{eq_bd}) into Eq.~(\ref{eq_condi}), we compute:
\begin{align}\label{eq_condi_calcu}
	&\left\langle\frac{\partial}{\partial\delta}(\mathcal{B}_1-\mathcal{D}_1)\right\rangle_{\begin{smallmatrix}\delta=0\\s_1=1\end{smallmatrix}}>0 \nonumber\\
    \Leftrightarrow&\left\langle\pi_1\right\rangle_{\begin{smallmatrix}\delta=0\\s_1=1\end{smallmatrix}}-\left\langle \sum_{j,l\in V}e_{1j}^{[R]}e_{jl}^{[R]}\pi_l\right\rangle_{\begin{smallmatrix}\delta=0\\s_1=1\end{smallmatrix}}>0 \nonumber\\
    \Leftrightarrow&~\pi^{(0,0)}-\pi^{(0,2)}>0\,.
\end{align}

Eq.~(\ref{eq_condi_calcu}) employs random walk notation. Specifically, an $(n,m)$-random walk involves $n$ steps on the interaction graph and $m$ steps on the learning graph. The expected value of a variable at the end of an $(n,m)$-random walk is denoted by $x^{(n,m)}$, where $x$ may represent $s$, $\pi$, or $F$. Since the walk occurs through edges, we obtain the following expression for the initial cooperator $1$:
\begin{equation}\label{eq_piwalk}
    \pi^{(0,0)}=\left\langle\pi_1\right\rangle_{\begin{smallmatrix}\delta=0\\s_1=1\end{smallmatrix}},~
    \pi^{(0,2)}=\left\langle \sum_{j,l\in V}e_{1j}^{[R]}e_{jl}^{[R]}\pi_l\right\rangle_{\begin{smallmatrix}\delta=0\\s_1=1\end{smallmatrix}}\,,
\end{equation}
which completes the final calculation step in Eq.~(\ref{eq_condi_calcu}).

The payoff calculation in Eq.~(\ref{eq_pii}) can also be straightforwardly rewritten using random walk terminology,
\begin{equation}\label{eq_pinm}
    \pi^{(n,m)}=-cs^{(n,m)}+bs^{(n+1,m)}\,.
\end{equation}

According to \cite{allen2014games,su2019spatial}, the following equation holds since we do not consider mutation:
\begin{equation}\label{eq_stop}
	s^{(n,m)}-s^{(n,m+1)}=\frac{\mu}{2}(Np^{(n,m)}-1)+\mathcal{O}(\mu^2)\,,
\end{equation}
where $\mu$ is an auxiliary parameter, which will be eliminated later, and $\mathcal{O}(\mu^2)\to 0$. Here, $p^{(n,m)}$ denotes the probability of ending at the starting vertex after the $(n,m)$-random walk. We will discuss the proper calculation of $p^{(n,m)}$ later.

Using Eq.~(\ref{eq_stop}), we construct the following equation:
\begin{align}\label{eq_stop_2}
	&s^{(n,m)}-s^{(n,m+2)} \nonumber\\
 =&~(s^{(n,m)}-s^{(n,m+1)})+(s^{(n,m+1)}-s^{(n,m+2)}) \nonumber\\
 =&~\frac{\mu}{2}(Np^{(n,m)}+Np^{(n,m+1)}-2)+\mathcal{O}(\mu^2)\,.
\end{align}

Next, we calculate the cooperation success condition given by Eq.~(\ref{eq_condi_calcu}),
\begin{align}\label{eq_condi_calcu_1}
    &~\pi^{(0,0)}-\pi^{(0,2)}>0 \nonumber\\
    \Leftrightarrow&~(-cs^{(0,0)}+bs^{(1,0)})-(-cs^{(0,2)}+bs^{(1,2)})>0 \nonumber\\
    \Leftrightarrow&-c(s^{(0,0)}-s^{(0,2)})+b(s^{(1,0)}-s^{(1,2)})>0 \nonumber\\
    \Leftrightarrow&-c(Np^{(0,0)}+Np^{(0,1)}-2) \nonumber\\
    &+b(Np^{(1,0)}+Np^{(1,1)}-2)>0\,,
\end{align}
which uses Eq.~(\ref{eq_pinm}) to replace $\pi^{(n,m)}$ with $s^{(n,m)}$ first, and then employs Eq.~(\ref{eq_stop_2}) to replace $s^{(n,m)}$ with $p^{(n,m)}$.

The actual forms of $p^{(n,m)}$ values highlight the difference between our model and the classic case where self-loops are excluded. Without walking, one stays in the starting vertex, hence $p^{(0,0)}=1$. One cannot leave and return to the starting vertex within one step in the classic case; however, in our model, one can walk to itself because self-loop is allowed with weight $w_R$ or $w_I$ on the learning and interaction graphs. Therefore, $p^{(0,1)}=w_R$ and $p^{(1,0)}=w_I$. Finally, there are two cases for $p^{(1,1)}$: one walks to itself in both steps, thus staying in the original place, with probability $w_Iw_R$, or, one walks to an arbitrary neighbor in the first step and goes back to the exact starting vertex in the second step, with probability $(1-w_I)(1-w_R)/k$. Therefore, $p^{(1,1)}=w_Iw_R+(1-w_I)(1-w_R)/k$.

Substituting the values of $p^{(0,0)}$, $p^{(0,1)}$, $p^{(1,0)}$, and $p^{(1,1)}$ into Eq.~(\ref{eq_condi_calcu_1}), we can finalize the calculation of cooperation success condition:
\begin{align}\label{eq_condi_calcu_2}
    &~\pi^{(0,0)}-\pi^{(0,2)}>0 \nonumber\\
    \Leftrightarrow&-c(N+Nw_R-2) \nonumber\\
    &+b\left[Nw_I+N\left(w_Iw_R+\frac{(1-w_I)(1-w_R)}{k}\right)-2\right]>0 \nonumber\\
    \Leftrightarrow&~\frac{b}{c}>\frac{N-2+N w_R}{N(k-1)w_I+N(k+1)w_I w_R+N-2k-Nw_R}k \nonumber\\
    &~~~\,\equiv \left(\frac{b}{c}\right)^\star\,.
\end{align}
Accordingly, when $b/c>(b/c)^\star$, cooperation is favored. The critical value $(b/c)^\star$ depends only on the population size $N$, the degree $k$ of vertices, and the weight factors $w_I$, $w_R$.

Furthermore, by combining our expression for $-c(Np^{(0,0)}+Np^{(0,1)}-2)+b(Np^{(1,0)}+Np^{(1,1)}-2)$ in Eq.~(\ref{eq_condi_calcu_2}) with the results obtained in Refs.~\cite{allen2017evolutionary,chen2013sharp} for the death-birth updating rules, we derive the theoretical cooperation level:
\begin{align}\label{eq_rhodg}
    \rho_C=&~\frac{N_C}{N}+\frac{N_C(N-N_C)}{2N(N-1)} \bigg\{-c(N+Nw_R-2) \nonumber\\
    &+b\left[Nw_I+N\left(w_Iw_R+\frac{(1-w_I)(1-w_R)}{k}\right)-2\right]\bigg\}\delta\,.
\end{align}
This term is a linear function of the benefit $b$ and cost $c$, with six other parameters: the selection strength $\delta$, the initial number of cooperators $N_C$, population $N$, degree $k$, self-weight for interaction $w_I$, and for updating $w_R$. Since Eq.~(\ref{eq_rhodg}) is linear, we should set $\rho_C=0$ if $\rho_C<0$ and $\rho_C=1$ if $\rho_C>1$ for self-consistency. It is important to note, however, that while the theoretical cooperation level predicted by Eq.~(\ref{eq_rhodg}) may approximate the results of MC simulations in a wide range, it is only strictly accurate as $\delta\to 0^+$ and $(b/c)\to (b/c)^\star$.

\subsection{The conflict between self-interaction and updating passivity}
Table~\ref{tablevalue} summarizes the main results concerning the threshold $(b/c)^\star$ for cooperation success, including the reduced form of $(b/c)^\star$ under specific parameters ($w_R=0$, $w_I=0$, $w_R=w_I\equiv w$) and the large population limit ($N\to +\infty$).

\begin{table*}[width=2\linewidth,pos=h]
\caption{Critical values of $b/c>(b/c)^\star$ for cooperation success under typical parameter values. All results are obtained by substituting specific parameter values into Eq.~(\ref{eq_condi_calcu_2}).}\label{tablevalue}
\begin{tabular*}{\tblwidth}{@{} LL@{} }
\toprule
\multicolumn{1}{L}{Special parameter} & 
The critical $(b/c)^\star$ for cooperation success
\\\midrule
/ &
$\displaystyle{\left(\frac{b}{c}\right)^\star=
\frac{N-2+N w_R}{N(k-1)w_I+N(k+1)w_I w_R+N-2k-Nw_R}k}$
\\
$w_I=0$ & 
$\displaystyle{\left(\frac{b}{c}\right)^\star=
\frac{N-2+N w_R}{N-2k-N w_R}k}$
\rule{0em}{2em}\\
$w_R=0$ &
$\displaystyle{\left(\frac{b}{c}\right)^\star=
\frac{N-2}{N(k-1)w_I+N-2k}k}$
\rule{0em}{2em}\\
$w_R=w_I\equiv w$ & 
$\displaystyle{\left(\frac{b}{c}\right)^\star=
\frac{N-2+Nw}{N-2k+N(k-2)w+N(k+1)w^2}k}$
\rule{0em}{2em}\\
$N\to +\infty$ & 
$\displaystyle{\left(\frac{b}{c}\right)^\star=
\frac{1+w_R}{(k-1)w_I+(k+1)w_I w_R+1-w_R}k}$
\rule{0em}{2em}\\
$N\to +\infty$, $w_I=0$ & 
$\displaystyle{\left(\frac{b}{c}\right)^\star=
\frac{1+w_R}{1-w_R}k}$
\rule{0em}{2em}\\
$N\to +\infty$, $w_R=0$ & 
$\displaystyle{\left(\frac{b}{c}\right)^\star=
\frac{1}{(k-1)w_I+1}k}$
\rule{0em}{2em}\\
$N\to +\infty$, $w_R=w_I\equiv w$ & 
$\displaystyle{\left(\frac{b}{c}\right)^\star=
\frac{1+w}{1+(k-2)w+(k+1)w^2}k}$
\rule{0em}{2em}\\
\bottomrule

\end{tabular*}
\end{table*}

\begin{figure*}
	\centering
		\includegraphics[width=.7\textwidth]{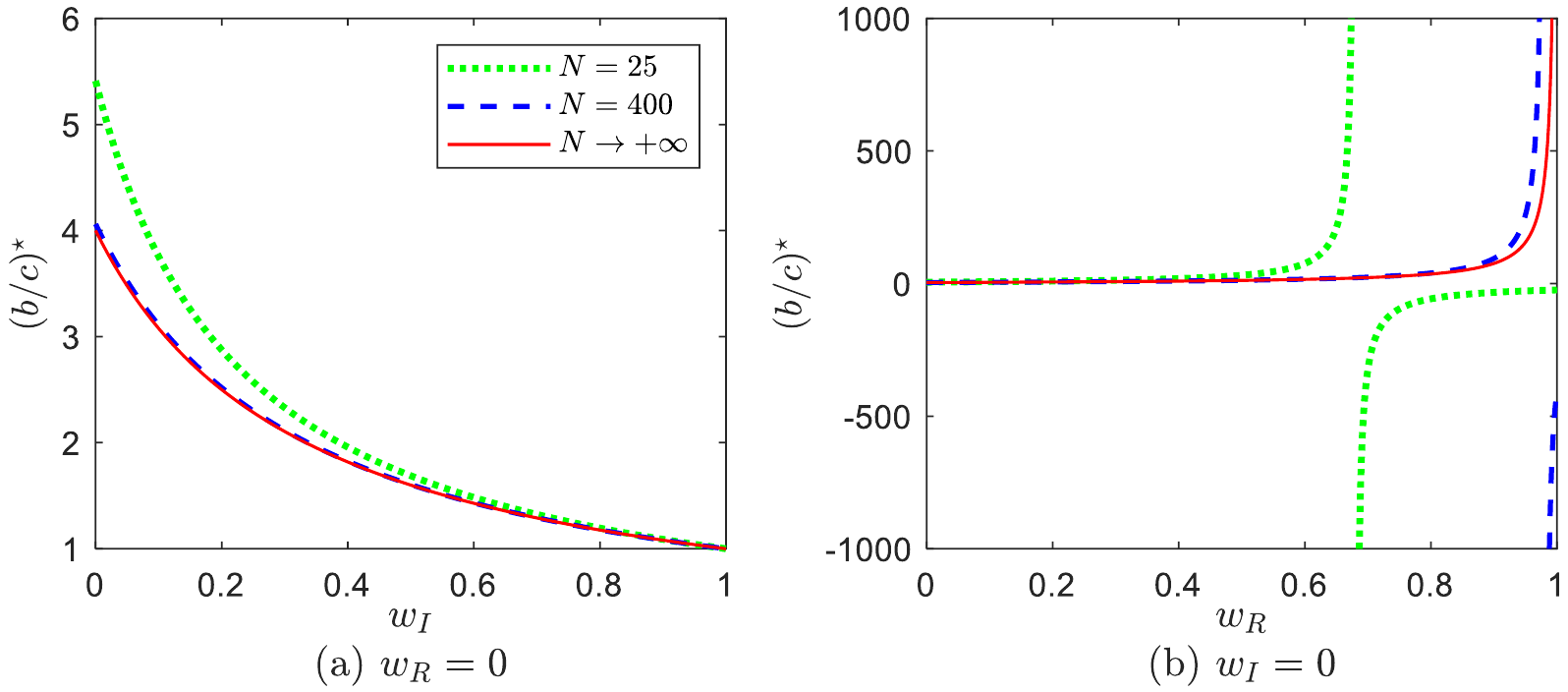}
	\caption{(a) The critical benefit-to-cost ratio $(b/c)^\star$ as a function of self-interaction weight $w_I$ when $w_R=0$. Self-interaction consistently promotes cooperation when acting alone. (b) The critical benefit-to-cost ratio $(b/c)^\star$ as a function of strategy updating inertia $w_R$ when $w_I=0$. Updating passivity impedes cooperation when acting alone. $k=4$.} 
	\label{fig_theowiwr}
\end{figure*}

From these results, we can make several observations. On the one hand, when $w_R=0$, the dependence of $(b/c)^\star$ on $w_I$ highlights the direct impact of ``self-gaming'': an increase in weight factor $w_I$ consistently reduces $(b/c)^\star$, thereby fostering cooperation. Fig.~\ref{fig_theowiwr}(a) demonstrates several representative cases for this function, and the effect of self-interaction is robust across different population sizes. Furthermore, in the $w_I \to 1$ limit, the cooperation success condition becomes $(b/c)^\star>1$ or $b>c$, resulting in a consistent preference for cooperation. On the other hand, when $w_I=0$, the dependence of $(b/c)^\star$ on $w_R$ echoes the system behavior previously reported in Ref.~\cite{wang2023evolution}: an increase in strategy updating inertia consistently raises $(b/c)^\star$, thus hindering cooperation. This effect is depicted in Fig.~\ref{fig_theowiwr}(b), where $(b/c)^\star$ is plotted as a function of $w_R$ at $w_I=0$ for various system sizes. A finite system size exhibits a unique behavior: beyond a certain $w_R$ threshold, $(b/c)^\star<0$ and cooperation success requires an unattainable $b/c<(b/c)^\star$ condition.

\begin{figure*}
	\centering
		\includegraphics[width=\textwidth]{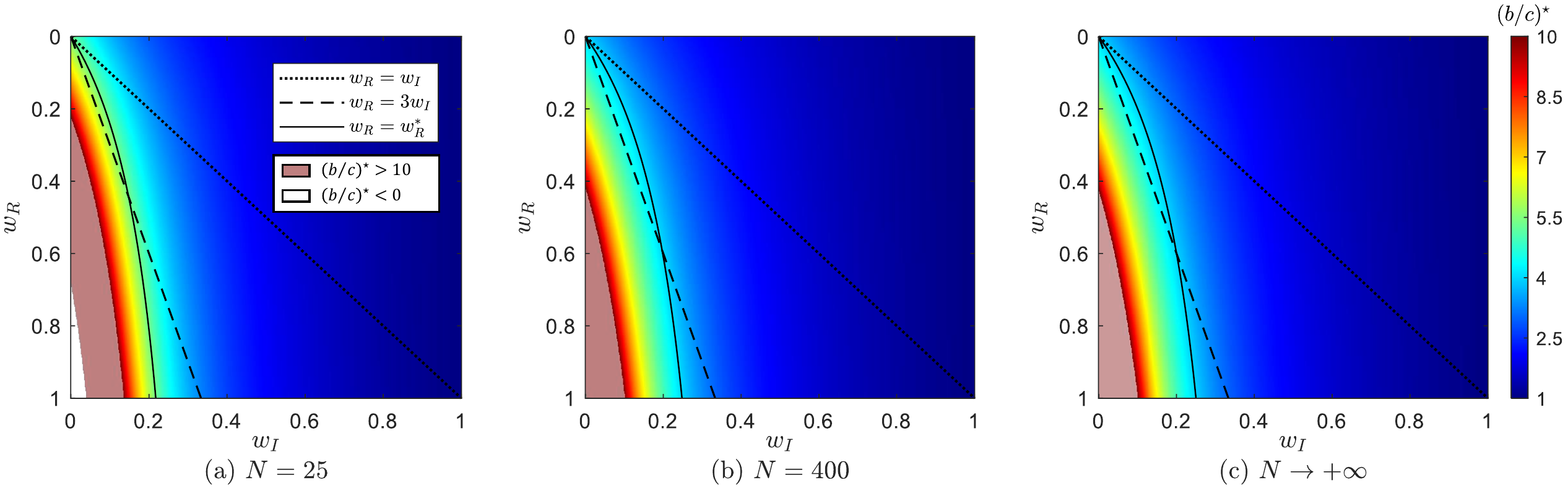}
	\caption{The critical benefit-to-cost ratio $(b/c)^\star$ as a bivariate function of self-interaction weight $w_I$ and self-learning weight $w_R$ for (a) $N=25$, (b) $N=400$, and (c) $N\to+\infty$. $k=4$.} 
	\label{fig_theobcmap}
\end{figure*}

These system behaviors elucidate the conflict between self-interaction and updating passivity. Our primary aim is to determine whether the positive influence of self-interaction or the negative effect of updating passivity prevails. To address this question, we first set $w_R=w_I\equiv w$ and analyze the impact of $w$, which intuitively signifies an equal weight of self-loops on both interaction and learning graphs. According to Table~\ref{tablevalue}, an increase in $w$ promotes cooperation. However, the overall system behavior is more intricate, as demonstrated when calculating the critical $(b/c)^\star$ value across the complete $w_R$-$w_I$ parameter plane. Fig.~\ref{fig_theobcmap} displays the full landscape of the critical benefit-to-cost ratio for small, practically large, and infinite system sizes. As a technical note, we have omitted details when $(b/c)^\star>10$ and $(b/c)^\star<0$ to ensure visibility. These ranges represent the parameter areas where achieving cooperation success is extremely challenging or impossible.

\begin{figure*}
	\centering
		\includegraphics[width=.95\textwidth]{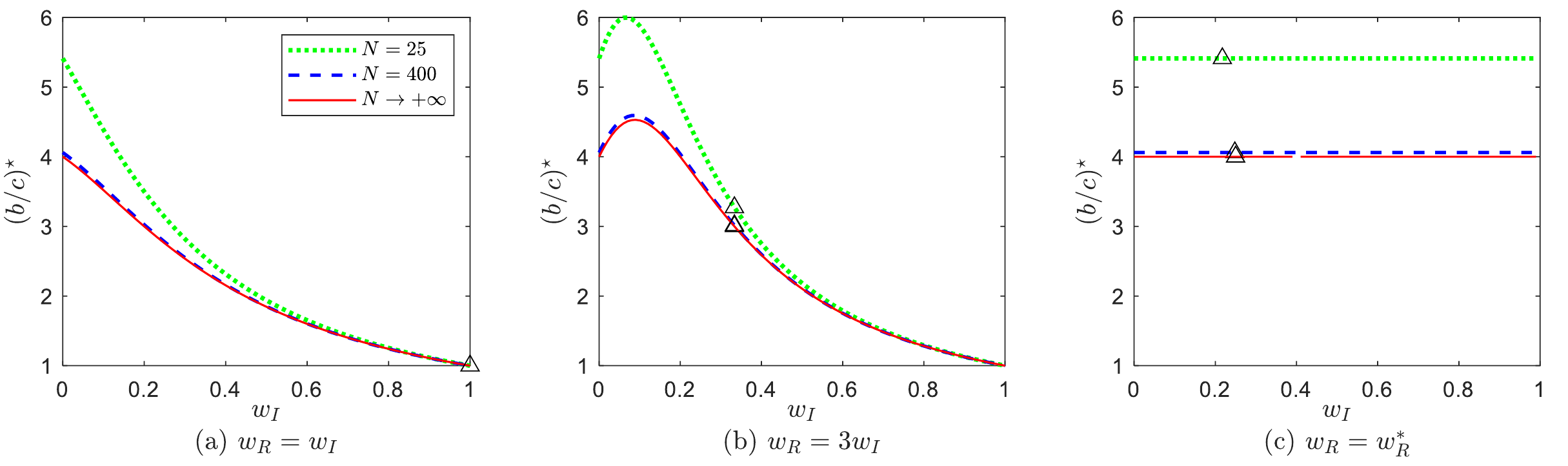}
	\caption{The critical benefit-to-cost ratio $(b/c)^\star$ as a function of self-interaction weight $w_I$ when $w_I$ and $w_R$ vary according to a specific trajectory: (a) $w_R=w_I$, (b) $w_R=3w_I$, or (c) $w_R=w_R^*$ as per Eq.~(\ref{eq_contourspecial}). Since $w_R<1$, the curves are meaningful when $w_I$ is on the left side of $\triangle$ where $w_R=1$. $k=4$.} 
	\label{fig_theowi}
\end{figure*}

As $w_I$ increases along the horizontal axis, the threshold value consistently declines, while the increase of $w_R$ along the vertical axis results in a larger $(b/c)^\star$. It is important to note that the origin is located at the upper-left corner of the parameter plane.
Though the effect of $w_I$ outweighs $w_R$ and moving along the $w_R=w_I$ line favors cooperation [see Fig.~\ref{fig_theowi}(a)], $w_R$ causes a more dramatic increase in $(b/c)^\star$. This increase is more pronounced for larger $w_R$ values compared to $w_I$, leading to a ``high hill'' in the region where $w_R$ is substantial and $w_I$ is small. This feature can result in some non-monotonic effects on the cooperation success threshold when traversing the surface. For instance, by adopting a simple linear relationship, $w_R=3w_I$, we can observe that an increase in $w_I$ initially hinders cooperation before promoting it, as illustrated in Fig.~\ref{fig_theowi}(b). As shown in Fig.~\ref{fig_theobcmap}(b), numerous trajectories of $w_R$ and $w_I$ on the $w_R$-$w_I$ plane can be detected where $(b/c)^\star$ increases and subsequently decreases with $w_I$ (first ascending and then descending the hill of large $w_R$ values). For simplicity, we will only present the relevant phenomenon using the straightforward linear constraint $w_R=3w_I$ for the remainder of this work.

Additionally, we can analytically determine the expression of the contour line of the ``hill'' at an arbitrary height $(b/c)^\star$. On the $w_R$-$w_I$ plane, we can identify a curve representing the same value of $(b/c)^\star$ by solving the expression of $(b/c)^\star$ in Table~\ref{tablevalue} and obtaining $w_R$ as a function of $w_I$, denoted by $w_R^*$,
\begin{equation}\label{eq_contourgeneral}
    w_R^*=\frac{N(k-1)(b/c)^\star w_I+(N-2k)(b/c)^\star-(N-2)k}{Nk+N(b/c)^\star -N(k+1)(b/c)^\star w_I}\,.
\end{equation}
In particular, the contour line maintaining the classic $(b/c)^\star$ value (by setting $w_R=0$ and $w_I=0$) can be obtained by substituting this expression of $(b/c)^\star$ into Eq.~(\ref{eq_contourgeneral}). In this case, we have
\begin{equation}\label{eq_contourspecial}
    w_R^*=\frac{(N-2)(k-1)w_I}{2(N-k-1)-(N-2)(k+1)w_I}\,,
\end{equation}
which represents the case we discuss for the remainder of this work. As shown in Fig.~\ref{fig_theowi}(c), along the constraint depicted by Eq.~(\ref{eq_contourspecial}), $(b/c)^\star$ remains invariant.

\section{Numerical simulation}\label{sec_nume}
In this section, we verify our theoretical conclusions through Monte Carlo simulations. Initially, we assign each agent a random strategy of cooperation or defection, resulting in an approximate initial number of cooperators $N_C\approx N/2$ and an initial cooperation level of $p_C(t_0)\approx 1/2$. As discussed at the beginning of Section~\ref{sec_theo}, evolution favors cooperation if $\rho_C>1/2$. We record the final cooperation level ($p_C=0$ or $p_C=1$) in the last MC step for each run. If the system does not reach fixation before the maximum step of $t=400000$~\cite{allen2017evolutionary}, we record the actual cooperation level ($0<p_C<1$) in the last MC step. The expected cooperation level $\rho_C$ under the given parameter values is obtained by averaging the outcomes of many independent runs. We investigate three representative population sizes: $5\times5$, $20\times20$, and $100\times100$ square lattices, where $N=25$, $400$, and $10,000$, respectively. Based on our empirical exploration of the system's relaxation, we set $\delta=0.01$ for the $5\times5$ and $20\times20$ lattices, averaging the outcomes of $10^6$ and $10^4$ runs, respectively, while for the $100\times100$ lattice, we set $\delta=0.1$ and record the outcome of a single run.

\begin{figure*}
	\centering
		\includegraphics[width=.98\textwidth]{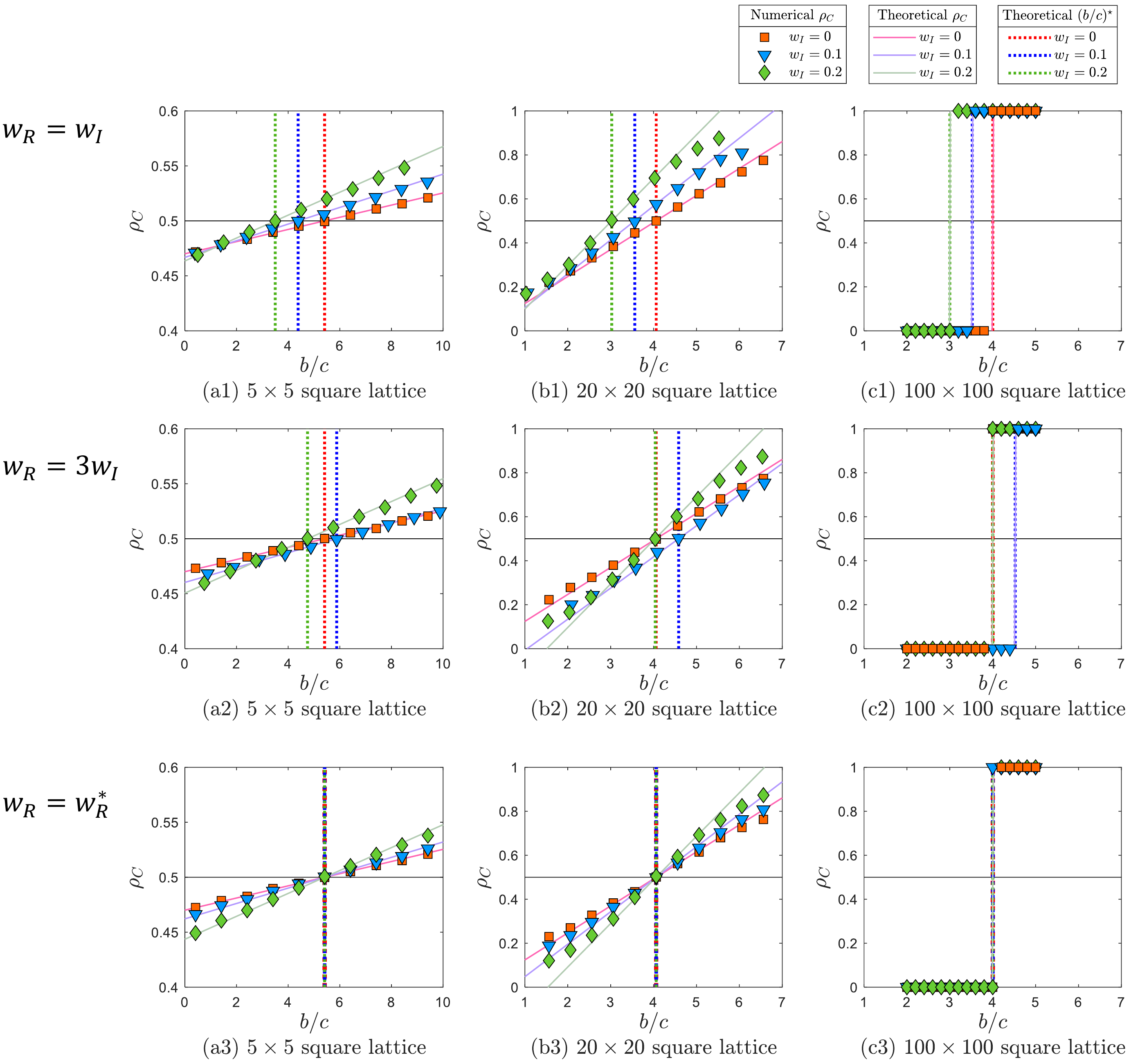}
	\caption{Monte Carlo simulations for the donation game (DG) corroborate the theoretical analysis. The numerical cooperation level $\rho_C$ is derived from the Monte Carlo simulation as detailed in Sections~\ref{sec_model} and \ref{sec_nume}, while the theoretical cooperation level $\rho_C$ is calculated using Eq.~(\ref{eq_rhodg}). The theoretical cooperation success threshold $(b/c)^\star$ is determined by Eq.~(\ref{eq_condi_calcu_2}). First row: the concurrent increase of weight factors promotes cooperation. Second row: when the self-learning weight increases more rapidly than self-interaction, the threshold level first rises and then falls. Third row: following a specific trajectory where $w_R=w_R^*$ according to Eq.~(\ref{eq_contourspecial}), the threshold level remains constant. Other parameters: $k=4$; $\delta=0.01$ for (a1)-(a3) and (b1)-(b3), $\delta=0.1$ for (c1)-(c3); $N_C=N/2$; $c=1$.} 
	\label{fig_dg}
\end{figure*}

Fig.~\ref{fig_dg} illustrates the results of Monte Carlo simulations for the donation game (DG) for the three trajectories discussed previously, with the horizontal axis representing a fixed value of $c=1$ while $b$ varies. In the first row, where $L=5$, the theoretical cooperation level is $\rho_C=\frac{1}{2}-\frac{23}{768}c+\frac{17}{3072}b$ when $w_I=0$ and the threshold for cooperation success is $(b/c)^\star\approx 5.4118$. If $w_I=0.1$, we have $\rho_C=\frac{1}{2}-\frac{17}{512}c+\frac{31}{4096}b$ and $(b/c)^\star\approx 4.3871$. Finally, when $w_I=0.2$, $\rho_C=\frac{1}{2}-\frac{7}{192}c+\frac{1}{96}b$ and $(b/c)^\star=3.5000$. In the second panel, where $L=20$, $\rho_C=\frac{1}{2}-\frac{199}{399}c+\frac{7}{57}b$ and $(b/c)^\star\approx 4.0612$ for $w_I=0$. If $w_I=0.1$, we have $\rho_C=\frac{1}{2}-\frac{73}{133}c+\frac{41}{266}b$ and $(b/c)^\star\approx 3.5610$. Lastly, at $w_I=0.2$, we have $\rho_C=\frac{1}{2}-\frac{239}{399}c+\frac{79}{399}b$ and $(b/c)^\star\approx 3.0253$. The third panel displays the results for the lattice with $L=100$. With $w_I=0$, we obtain $\rho_C=\frac{1}{2}-\frac{1249750}{9999}c+\frac{312250}{9999}b$ and $(b/c)^\star\approx 4.0024$. For $w_I=0.1$, the results yield $\rho_C=\frac{1}{2}-\frac{152750}{1111}c+\frac{43375}{1111}b$ and $(b/c)^\star\approx 3.5216$. Finally, at $w_I=0.2$, we have $\rho_C=\frac{1}{2}-\frac{1499750}{9999}c+\frac{499750}{9999}b$ and $(b/c)^\star\approx 3.0010$. The comparison of different $(b/c)^\star$ values confirms that the simultaneous increase of weight factors promotes cooperation.

The second row of Fig.~\ref{fig_dg} illustrates the scenario where we follow the $w_R=3w_I$ trajectory. For $L=5$, when $w_I=0$, we obtain $\rho_C=\frac{1}{2}-\frac{23}{768}c+\frac{17}{3072}b$ and $(b/c)^\star\approx 5.4118$. With $w_I=0.1$, we have $\rho_C=\frac{1}{2}-\frac{61}{1536}c+\frac{83}{12288}b$ and $(b/c)^\star\approx 5.8795$. Lastly, for $w_I=0.2$, we find $\rho_C=\frac{1}{2}-\frac{19}{384}c+\frac{1}{96}b$ and $(b/c)^\star=4.7500$. When $L=20$ and $w_I=0$, $\rho_C=\frac{1}{2}-\frac{199}{399}c+\frac{7}{57}b$ and $(b/c)^\star\approx 4.0612$. Here, $w_I=0.1$ results in $\rho_C=\frac{1}{2}-\frac{37}{57}c+\frac{113}{798}b$ and $(b/c)^\star\approx 4.5841$. Finally, at $w_I=0.2$, we have $\rho_C=\frac{1}{2}-\frac{319}{399}c+\frac{79}{399}b$ and $(b/c)^\star\approx 4.0380$. For $L=100$ and $w_I=0$, we obtain $\rho_C=\frac{1}{2}-\frac{1249750}{9999}c+\frac{312250}{9999}b$ and $(b/c)^\star\approx 4.0024$. At $w_I=0.1$, $\rho_C=\frac{1}{2}-\frac{1624750}{9999}c+\frac{359125}{9999}b$ and $(b/c)^\star\approx 4.5242$. If $w_I=0.2$, we derive $\rho_C=\frac{1}{2}-\frac{1999750}{9999}c+\frac{499750}{9999}b$ and $(b/c)^\star\approx 4.0015$. In this case, when the self-learning weight factor increases more rapidly than the self-interaction weight, we observe a non-monotonous shift in the threshold values. Initially, the increase in weight factors inhibits cooperation, but later encourages it.

The third row of Fig.~\ref{fig_dg} presents the case when $w_R=w_R^*$ is maintained according to Eq.~(\ref{eq_contourspecial}). For $L=5$ and $w_I=0$, we obtain $\rho_C=\frac{1}{2}-\frac{23}{768}c+\frac{17}{3072}b$ and $(b/c)^\star\approx 5.4118$. At $w_I=0.1$, we have $\rho_C=\frac{1}{2}-\frac{23}{608}c+\frac{17}{2432}b$ and $(b/c)^\star\approx 5.4118$. For $w_I=0.2$, we find $\rho_C=\frac{1}{2}-\frac{23}{408}c+\frac{1}{96}b$ and $(b/c)^\star\approx 5.4118$. If $L=20$ and $w_I=0$, we have $\rho_C=\frac{1}{2}-\frac{199}{399}c+\frac{7}{57}b$ and $(b/c)^\star\approx 4.0612$. At $w_I=0.1$, we find $\rho_C=\frac{1}{2}-\frac{15721}{26201}c+\frac{553}{3743}b$ and $(b/c)^\star\approx 4.0612$. Lastly, at $w_I=0.2$, we have $\rho_C=\frac{1}{2}-\frac{15721}{19551}c+\frac{79}{399}b$ and $(b/c)^\star\approx 4.0612$. For $L=100$ and $w_I=0$, we obtain $\rho_C=\frac{1}{2}-\frac{1249750}{9999}c+\frac{312250}{9999}b$ and $(b/c)^\star\approx 4.0024$. At $w_I=0.1$, we find $\rho_C=\frac{1}{2}-\frac{412275645232534375}{2748504191533056}c+\frac{824056460724841625}{21988033532264448}b$ and $(b/c)^\star\approx 4.0024$. Finally, at $w_I=0.2$, we obtain $\rho_C=\frac{1}{2}-\frac{12495702011469625}{62466004353024}c+\frac{499750}{9999}b$ and $(b/c)^\star\approx 4.0024$. In conclusion, since $(b/c)^\star$ values remain invariant, the increase of weight factors does not influence the threshold level for cooperation success. However, it is evident that the slope of the theoretical cooperation level $\rho_C$ increases with a larger weight factor.

\section{Extension to alternative games}
To examine the robustness of our findings, we can extend the model to different types of games. In an arbitrary two-player game, the strategy of agent $i$ can be denoted by a vector $\mathbf{s}_i=(s_i,1-s_i)^\mathrm{T}$, where $\mathbf{s}_i=(1,0)^\mathrm{T}$ represents cooperation and $\mathbf{s}_i=(0,1)^\mathrm{T}$ denotes defection. Similar to Eq.~(\ref{eq_pii}), the payoff that agent $i$ receives through the interaction graph can be calculated using:
\begin{equation}\label{eq_pii_arbitrary}
    \pi_i=\sum_{l\in V}(e_{il}^{[I]}\mathbf{s}_i^\mathrm{T}\cdot\mathbf{M}\cdot\mathbf{s}_l)\,,
\end{equation}
where $\mathbf{M}$ is a $2\times 2$ payoff matrix, 
\begin{equation}
    \mathbf{M}=\begin{pmatrix}
    \mathbb{R} & \mathbb{S}\\
    \mathbb{T} & \mathbb{P}
    \end{pmatrix}\,.
\end{equation}
In conventional notation, $\mathbb{R}$ denotes the reward for mutual cooperation, $\mathbb{T}$ signifies the temptation to defect, $\mathbb{S}$ represents the sucker's payoff, and $\mathbb{P}$ indicates the punishment for defection.

The arbitrary game in Eq.~(\ref{eq_pii_arbitrary}) reduces to the donation game in Eq.~(\ref{eq_pii}) when $\mathbf{M}$ takes the following form:
\begin{equation}\label{eq_Mdg}
    \mathbf{M}=\begin{pmatrix}
    b-c & -c\\
    b & 0
    \end{pmatrix}\,.
\end{equation}

According to the structure coefficient theorem~\cite{tarnita2009strategy}, the condition for the success of cooperation in a general two-player game can be expressed as:
\begin{equation}\label{eq_sigmatheorem}
    \sigma\mathbb{R}+\mathbb{S}>\mathbb{T}+\sigma\mathbb{P}\,,
\end{equation}
where $\sigma$ is the structure coefficient independent of the payoff matrix. We can determine the $\sigma$ value here by the results of the donation game. By applying the payoff matrix in Eq.~(\ref{eq_Mdg}) to Eq.~(\ref{eq_sigmatheorem}) and comparing it with the cooperation success condition of the donation game using Eq.~(\ref{eq_condi_calcu_2}), we obtain:
\begin{equation}
    \sigma=\frac{(b/c)^\star+1}{(b/c)^\star-1}\,,
\end{equation}
where $(b/c)^\star$ is provided by Eq.~(\ref{eq_condi_calcu_2}).

Furthermore, we can derive the theoretical cooperation level $\rho_C$ from the structure coefficient theorem and the expression of $\rho_C$ for the donation game. The core idea is to construct the donation game and arbitrary games in the same form using the structure coefficient theorem in Eq.~(\ref{eq_sigmatheorem}), without multiplying or dividing by any quantity in the process. Then, substitute the part of the arbitrary games in the structure coefficient theorem that is equal to the donation game back into Eq.~(\ref{eq_rhodg}).

By applying $\sigma=[(b/c)^\star+1]/[(b/c)^\star-1]$ to $\sigma (b-c)-c>b\Leftrightarrow -(\sigma+1)c+(\sigma-1)b>0$, we find that $-(\sigma+1)c+(\sigma-1)b$ is multiplied by $2k/\{(N-2+Nw_R)k-[N(k-1) w_I+N(k+1) w_I w_R+N-2k-Nw_R ]\}$ based on $-c(Np^{(0,0)}+Np^{(0,1)}-2)+b(Np^{(1,0)}+Np^{(1,1)}-2)$. The same part of arbitrary games $\sigma\mathbb{R}+\mathbb{S}>\mathbb{T}+\sigma\mathbb{P}\Leftrightarrow \sigma(\mathbb{R}-\mathbb{P})+(\mathbb{S}-\mathbb{T})>0$ is $\sigma(\mathbb{R}-\mathbb{P})+(\mathbb{S}-\mathbb{T})$, and we divide it by $2k/\{(N-2+Nw_R)k-[N(k-1) w_I+N(k+1) w_I w_R+N-2k-Nw_R ]\}$, substituting it back to the same position in Eq.~(\ref{eq_rhodg}). In this manner, we obtain the theoretical cooperation level $\rho_C$ for arbitrary two-player games as follows:
\begin{align}\label{eq_rhoarbitrary}
    \rho_C=&~\frac{N_C}{N}+\frac{N_C(N-N_C)}{4N(N-1)}
    \bigg\{ (N-2+Nw_R) \nonumber\\
    &\times (\mathbb{R}+\mathbb{S}-\mathbb{T}-\mathbb{P}) \nonumber\\
    & +\frac{N(k-1) w_I+N(k+1) w_I w_R+N-2k-Nw_R}{k} \nonumber\\
    &\times (\mathbb{R}-\mathbb{S}+\mathbb{T}-\mathbb{P})\bigg\}\delta\,, 
\end{align}
which is a function of $\mathbb{R}$, $\mathbb{S}$, $\mathbb{T}$, and $\mathbb{P}$, with six other parameters as mentioned below Eq.~(\ref{eq_rhodg}).
It is essential to clarify that our approach for deducing $\rho_C$ for arbitrary two-player games above is non-rigorous, although it may predict Monte Carlo simulations, as we will see later.

In the following, we apply the general results to three representative social games, which include the prisoner's dilemma game, the stag-hunt game, and the snowdrift game.

\subsection{The prisoner's dilemma game}
For simplicity, we consider the so-called weak prisoner's dilemma game (PD)~\cite{nowak1992evolutionary,nowak1993spatial}, where the temptation is the only control parameter. The payoff matrix is:
\begin{equation}\label{eq_Mpd}
    \mathbf{M}=\begin{pmatrix}
    1 & 0\\
    b_\text{PD} & 0
    \end{pmatrix}\,.
\end{equation}

According to Eq.~(\ref{eq_sigmatheorem}), the threshold of cooperation success $b_\text{PD}^\star$ is
\begin{equation}\label{eq_condipd}
\begin{split}
    b_\text{PD}^\star&=\sigma \\
    &=\frac{(k+1)-\frac{4k}{N}+(k-1) w_I+(k+1) w_I w_R+(k-1) w_R}{(k-1)-(k-1) w_I-(k+1) w_I w_R+(k+1) w_R}  
\end{split}
\end{equation}
and evolution favors cooperation if $b_\text{PD}<b_\text{PD}^\star$. Moreover, we have the theoretical cooperation level $\rho_C$ by substituting Eq.~(\ref{eq_Mpd}) into Eq.~(\ref{eq_rhoarbitrary}):
\begin{align}\label{eq_rhopd}
    \rho_C=&~\frac{N_C}{N}+\frac{N_C(N-N_C)}{4N(N-1)}
    \bigg\{ (N-2+Nw_R) \nonumber\\
    &\times (1-b_\text{PD}) \nonumber\\
    & +\frac{N(k-1) w_I+N(k+1) w_I w_R+N-2k-Nw_R}{k} \nonumber\\
    &\times (1+b_\text{PD})\bigg\}\delta\,, 
\end{align}
which is a function of $b_\text{PD}$, with six other parameters.

\begin{figure*}
	\centering
		\includegraphics[width=.9\textwidth]{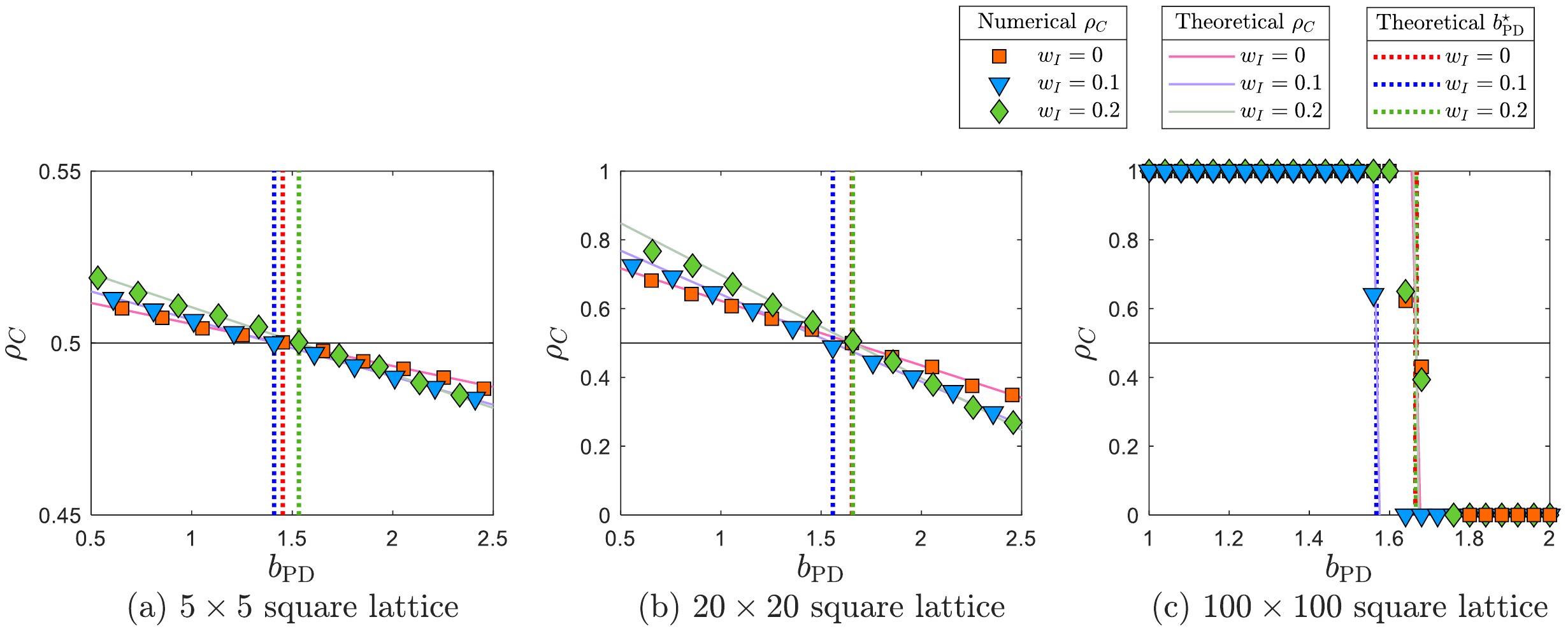}
	\caption{Monte Carlo simulations for the prisoner's dilemma game (PD) are well predicted by theoretical analysis. Numerical cooperation level $\rho_C$ is obtained by the Monte Carlo simulation with payoff matrix $\mathbf{M}$ in Eq.~(\ref{eq_Mpd}). Theoretical cooperation level $\rho_C$ is obtained by Eq.~(\ref{eq_rhopd}). Theoretical cooperation success threshold $b_\text{PD}^\star$ is obtained by Eq.~(\ref{eq_condipd}). $w_R=3w_I$, and other parameters are the same as the ones in the simulation of DG.}\label{fig_pd}
\end{figure*}

Fig.~\ref{fig_pd} shows the cooperation level $\rho_C$ as a function of temptation when $w_R=3w_I$, which provides the most illustrative example of the conflict between self-interaction and updating passivity. We can see that $\rho_C$ obtained by Monte Carlo simulations agrees with the prediction of theoretical $\rho_C$ calculated by Eq.~(\ref{eq_rhopd}). Meanwhile, from $w_I=0$ to $w_I=0.1$ and $w_I=0.2$, the threshold $b_\text{PD}^\star$ for cooperation success first decreases and then increases, which means the increase of weight factors first inhibits and later promotes cooperation.

\subsection{The stag-hunt game}
The stag-hunt game (SH), as delineated by previous studies~\cite{starnini2011coordination,wang2013evolving,dong2019memory}, employs a single-parameter payoff matrix of $r_\text{SH}$, presented below.
 \begin{equation}\label{eq_Msh}
    \mathbf{M}=\begin{pmatrix}
    1 & -r_\text{SH}\\
    r_\text{SH} & 0
    \end{pmatrix}\,.
\end{equation}

Drawing from Eq.~(\ref{eq_sigmatheorem}), the cooperation success threshold $r_\text{SH}^\star$ is given by
\begin{align}\label{eq_condish}
    r_\text{SH}^\star&=\frac{\sigma}{2} \nonumber\\
    &=\frac{(k+1)-\frac{4k}{N}+(k-1) w_I+(k+1) w_I w_R+(k-1) w_R}{2\left[(k-1)-(k-1) w_I-(k+1) w_I w_R+(k+1) w_R\right]}\,.
\end{align}
Evolution favors cooperation when $r<r_\text{SH}$. Furthermore, the theoretical cooperation level, $\rho_C$, can be expressed as
\begin{align}\label{eq_rhosh}
    \rho_C=&~\frac{N_C}{N}+\frac{N_C(N-N_C)}{4N(N-1)}
    \bigg\{ (N-2+Nw_R) \nonumber\\
    &\times (1-2r_\text{SH}) \nonumber\\
    & +\frac{N(k-1) w_I+N(k+1) w_I w_R+N-2k-Nw_R}{k} \nonumber\\
    &\times (1+2r_\text{SH})\bigg\}\delta\,, 
\end{align}
This formulation of $\rho_C$ is a function of $r_\text{SH}$, incorporating six other parameters.

\begin{figure*}
	\centering
		\includegraphics[width=.9\textwidth]{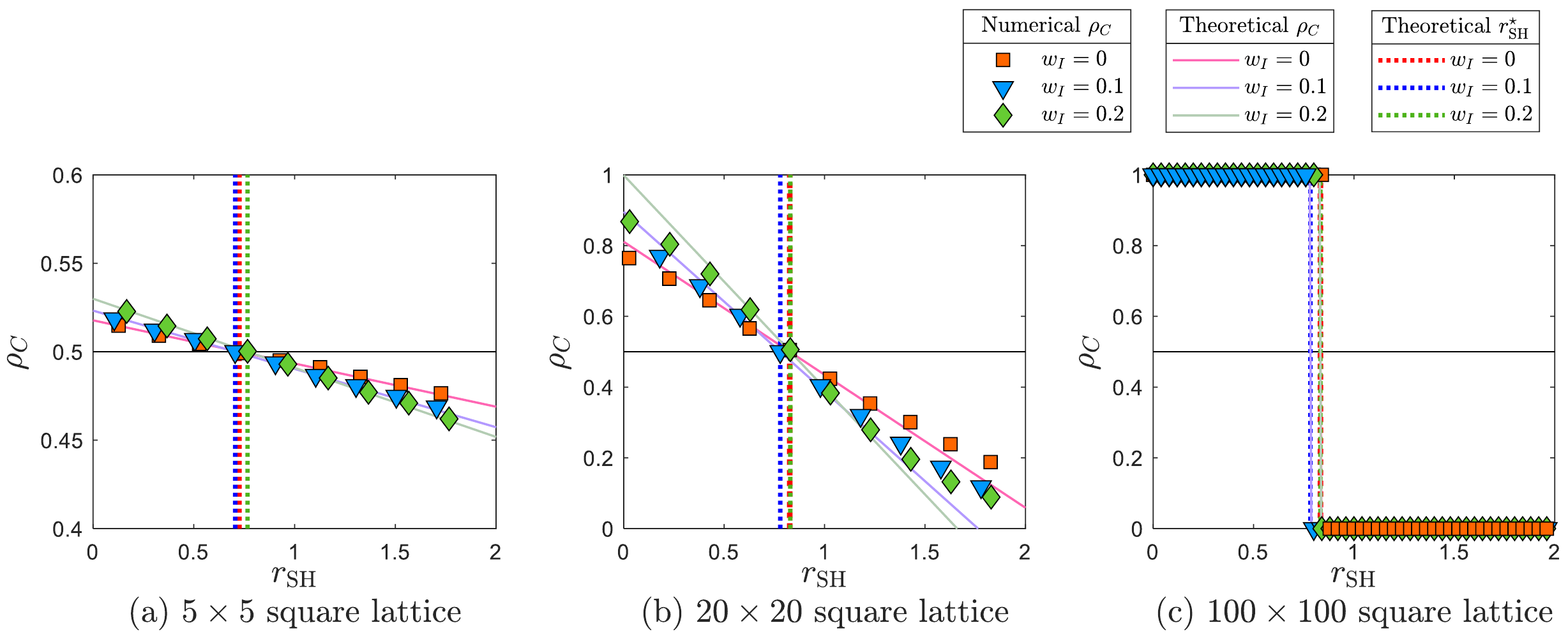}
	\caption{Monte Carlo simulations for the stag-hunt game (SH) align with theoretical analysis predictions. The numerical cooperation level $\rho_C$ is determined through Monte Carlo simulations, utilizing the payoff matrix $\mathbf{M}$ in Eq.~(\ref{eq_Msh}). The theoretical cooperation level $\rho_C$ is calculated via Eq.~(\ref{eq_rhosh}). The theoretical cooperation success threshold, $r_\text{SH}^\star$, is obtained from Eq.~(\ref{eq_condish}). The parameter $w_R=3w_I$, and other parameters correspond to those employed in the simulation of DG.} \label{fig_sh}
\end{figure*}

In a similar vein, Fig.~\ref{fig_sh} illustrates the cooperation level $\rho_C$ as a function of $r_\text{SH}$, given $w_R=3w_I$. The numerical $\rho_C$ derived from Monte Carlo simulations aligns with the theoretical prediction computed via Eq.~(\ref{eq_rhopd}). As before, the non-monotonic variation in the threshold value is observable when altering $w_I$ from 0 to $w_I=0.1$ and $w_I=0.2$.

\subsection{The snowdrift game}
The snowdrift game (SD)~\cite{hauert2004spatial,wang2006memory,zhang2012novel,su2017spatial,shu2018memory} also features a single-parameter payoff matrix of $r_\text{SD}$:
 \begin{equation}\label{eq_Msd}
    \mathbf{M}=\begin{pmatrix}
    1 & 1-r_\text{SD}\\
    1+r_\text{SD} & 0
    \end{pmatrix}\,.
\end{equation}

In accordance with Eq.~(\ref{eq_sigmatheorem}), the cooperation success threshold, $r_\text{SD}^\star$, is
\begin{align}\label{eq_condisd}
    r_\text{SD}^\star&=\frac{\sigma}{2} \nonumber\\
    &=\frac{(k+1)-\frac{4k}{N}+(k-1) w_I+(k+1) w_I w_R+(k-1) w_R}{2\left[(k-1)-(k-1) w_I-(k+1) w_I w_R+(k+1) w_R\right]}\,.
\end{align}
Cooperation is favored by evolution when $r<r_\text{SD}^\star$. The distinction in $\mathbb{S}$ and $\mathbb{T}$ between stag-hunt and snowdrift games vanishes within the structure coefficient theorem, Eq.~(\ref{eq_sigmatheorem}), equating the critical cooperation success thresholds and theoretical cooperation levels,
\begin{align}\label{eq_rhosd}
    \rho_C=&~\frac{N_C}{N}+\frac{N_C(N-N_C)}{4N(N-1)}
    \bigg\{ (N-2+Nw_R) \nonumber\\
    &\times (1-2r_\text{SD}) \nonumber\\
    & +\frac{N(k-1) w_I+N(k+1) w_I w_R+N-2k-Nw_R}{k} \nonumber\\
    &\times (1+2r_\text{SD})\bigg\}\delta\,, 
\end{align}

\begin{figure*}
	\centering
		\includegraphics[width=.9\textwidth]{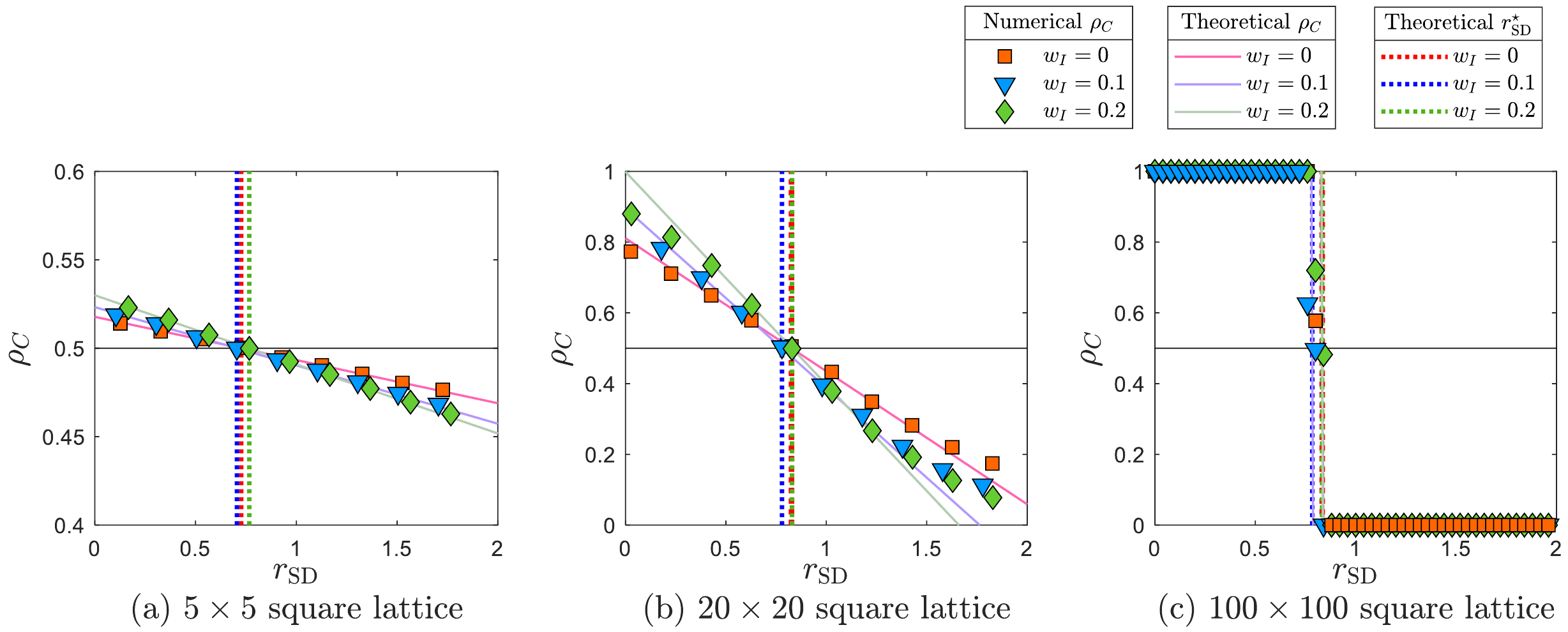}
	\caption{Monte Carlo simulations for the snowdrift game (SD) accord with the theoretical analysis. The numerical cooperation level $\rho_C$ is derived from the Monte Carlo simulations, employing the payoff matrix $\mathbf{M}$ in Eq.~(\ref{eq_Msd}). The theoretical cooperation level $\rho_C$ is ascertained using Eq.~(\ref{eq_rhosd}). The theoretical cooperation success threshold, $r_\text{SD}^\star$, is obtained from Eq.~(\ref{eq_condisd}). The parameter $w_R=3w_I$, and other parameters are consistent with those used in the simulation of DG.} \label{fig_sd}
\end{figure*}

Fig.~\ref{fig_sd} displays the cooperation level $\rho_C$ as a function of $r_\text{SD}$, with $w_R=3w_I$. The cooperation success threshold $r_\text{SD}^\star$ initially diminishes and subsequently escalates, mirroring the findings in the stag-hunt game context.

Comparing the prisoner's dilemma, the stag-hunt, and the snowdrift games, the most striking effect of $w_I$ can be detected in the prisoner's dilemma.

\section{Conclusion}
Prosocial behavior, such as donating a minor cost to provide a significant benefit to another individual, may appear contradictory to human beings' inherently selfish nature. Intuitively, if self-loop interactions are considered, where individuals can donate a small cost to themselves and directly receive the large benefit, cooperation would emerge without dependence on external conditions, such as network structure and updating rules. Meanwhile, previous research has demonstrated that increasing the self-loop in the updating process hinders cooperation~\cite{wang2023evolution}. This poses a potential paradox: does the self-loop in evolutionary game dynamics result in the positive effect of self-interaction or the negative effect of updating passivity?

The answer is nuanced, as the final effect may rely on the contributions of the aforementioned factors. Consequently, in this study, we introduced self-weights for playing games and updating strategies to characterize self-interaction and updating passivity. We analyzed the basic social dilemma, the donation game, on a square lattice and derived theoretical solutions for the cooperation success condition and cooperation level under weak selection. Our initial findings confirm that self-interaction consistently fosters cooperation, whereas updating passivity persistently inhibits cooperation. Building upon this, we discovered that the positive effect of self-interaction on cooperation surpasses the inhibitory influence of updating passivity, indicating that an equal increase in self-loop within evolutionary game dynamics indeed promotes cooperation.

Nevertheless, updating passivity can exert a substantial inhibitory effect on cooperation, although such severe inhibition occurs when updating passivity is large and self-interaction is small. We can derive constant cooperation success threshold contours on the $w_R$-$w_I$ plane. This suggests that even along a simple trajectory, for example, $w_R=3w_I$, self-loops may have a non-monotonic impact on cooperation. We observed that under the aforementioned constraint (i.e., when updating passivity is triple the self-interaction), cooperation is initially impeded but subsequently facilitated as self-interaction increases.

Moreover, we generalized our findings to encompass diverse games, examining three classic examples: the prisoner's dilemma, stag-hunt, and snowdrift. Our conclusions remain consistent across these different game scenarios. Future research could extend these conclusions to arbitrary network structures~\cite{allen2017evolutionary}, further broadening the understanding of cooperation dynamics on general social structures.

\printcredits

\section*{Acknowledgement}
A.S. was supported by the National Research, Development and Innovation Office (NKFIH) under Grant No. K142948.

\onecolumn
\appendix
\renewcommand\thefigure{\Alph{section}\arabic{figure}}
\renewcommand{\theequation}{\thesection.\arabic{equation}}

\section{Pair approximation}\label{sec_appen_pair}
\setcounter{figure}{0}
\setcounter{equation}{0}
To broaden our theoretical study and evaluate the robustness of our findings, we employ the pair approximation method~\cite{ohtsuki2006simple,li2014cooperation,su2019evolutionary}. As a type of mean-field approximation, the pair approximation belongs to a distinct technique family compared to the IBD method. Notably, pair approximation investigates dynamics within an infinite population. We will demonstrate that the same result as $N\to +\infty$, obtained using the IBD method, can also be acquired through pair approximation.

We summarize several useful equations derived from binomial theory for swift application in subsequent calculations. Assuming $k$ is a positive integer, $k_C$ is an integer between $0$ and $k$, $0\leq k_C\leq k$, and $z$ is a proportional quantity, $0\leq z\leq 1$, we have:
\begin{subequations}
\begin{align}
    &\sum_{k_C=0}^k{\frac{k!}{k_C!(k-k_C)!}z^{k_C}(1-z)^{k-k_C}}=1, \label{bino1}\\
    &\sum_{k_C=0}^k{\frac{k!}{k_C!(k-k_C)!}z^{k_C}(1-z)^{k-k_C}k_C}=kz, \label{bino2}\\
    &\sum_{k_C=0}^k{\frac{k!}{k_C!(k-k_C)!}z^{k_C}(1-z)^{k-k_C}{k_C}^2}=kz[1+(k-1)z], \label{bino3}\\
    &\sum_{k_C=0}^k{\frac{k!}{k_C!(k-k_C)!}z^{k_C}(1-z)^{k-k_C}k_C(k-k_C)}=k(k-1)z(1-z).\label{bino4}
\end{align}
\end{subequations}

\subsection{Constructing the system}
We denote the proportion of $C$-players and $D$-players in the system as $p_C$ and $p_D$, respectively. The probability of finding a $C$-player or $D$-player in the neighborhood of an $X$-player is denoted by $q_{C|X}$ and $q_{D|X}$, where $X$ represents either $C$ or $D$. The proportion of edges connecting a pair of $X$- and $Y$-players is denoted as $p_{XY}$, where $X$ and $Y$ may represent $C$, $D$. Due to constraints, their relations are as follows:
\begin{align}\label{eq_system}
    p_C+p_D&=1, \nonumber\\
    q_{C|X}+q_{D|X}&=1, \nonumber\\
    p_{XY}&=q_{X|Y} p_Y, \nonumber\\
    p_{CD}&=p_{DC}.
\end{align}
In total, we have nine variables: $p_C$, $p_D$, $q_{C|C}$, $q_{C|D}$, $q_{D|C}$, $q_{D|D}$, $p_{CC}$, $p_{CD}$, and $p_{DD}$. According to Eq.~(\ref{eq_system}), the system can be described using only two independent variables: $p_C$ and $q_{C|C}$. The remaining seven variables can be expressed as functions of $p_C$ and $q_{C|C}$ as follows:
\begin{align}\label{eq_otherv}
    p_{CC}&=q_{C|C} p_C, \nonumber\\
    p_D&=1-p_C, \nonumber\\
    q_{D|C}&=1-q_{C|C}, \nonumber\\
    p_{CD}&=p_C p_{D|C}=p_C(1-q_{C|C}), \nonumber\\
    q_{C|D}&=\frac{p_{CD}}{p_D}=\frac{p_C(1-q_{C|C})}{1-p_C}, \nonumber\\
    q_{D|D}&=1-q_{C|D}=\frac{1-2p_C+p_Cq_{C|C}}{1-p_C}, \nonumber\\
    p_{DD}&=p_Dq_{D|D}=1-2p_C+p_Cq_{C|C}.
\end{align}

\subsection{Updating a $D$-player}
First, we examine the case where the focal agent is a $D$-player. Let there be $k_C$ cooperators surrounding this $D$-player. In this scenario, the $D$-player's payoff is given by:
\begin{equation}\label{eq_pid}
    \pi_D=\frac{1-w_I}{k}k_C b=(1-w_I)\frac{k_C}{k}b.
\end{equation}

The expected payoff for a $C$-player around this focal $D$-player is
\begin{align}\label{eq_picd}
    \pi_{C|D}=&~w_I (-c+b)+\frac{1-w_I}{k}
    \left\{
    -c+\sum_{k_C'=0}^{k-1}{\frac{(k-1)!}{k_C'!(k-k_C'-1)!}q_{C|C}^{k_C'} q_{D|C}^{k-k_C'-1}[-(k-1)c+k_C' b]}
    \right\} \nonumber\\
    =&-c+w_I b+\frac{1-w_I}{k}(k-1)q_{C|C}b,
\end{align}
where the $C$-player plays the game with itself with weight $w_I$. Moreover, it pays a cost $c$ to the focal $D$-player and the remaining $k-1$ neighbors,  receiving $b$ from $k_C'$ cooperators among the $k-1$ remaining neighbors. The summation is computed using Eqs.~(\ref{bino1}) and (\ref{bino2}).

Likewise, the expected payoff for a $D$-player surrounding the focal $D$-player is
\begin{align}\label{eq_pidd}
    \pi_{D|D}=&~\frac{1-w_I}{k}
    \sum_{k_C'=0}^{k-1}{\frac{(k-1)!}{k_C'!(k-k_C'-1)!}q_{C|D}^{k_C'} q_{D|D}^{k-k_C'-1}k_C' b} \nonumber\\
    =&~\frac{1-w_I}{k}(k-1)q_{C|D}b.
\end{align}

We proceed to convert the payoff to fitness $F = \exp(\delta \pi)$. According to the death-birth process, the focal $D$-player becomes a $C$-player with probability
\begin{align}\label{eq_DgetsC}
    \mathcal{P}(D\gets C)=&~\frac{(1-w_R)/k\cdot k_C F_{C|D}}
    {w_R F_D+(1-w_R)/k\cdot [k_C F_{C|D}+(k-k_C)F_{D|D}]} \nonumber\\
    =&~(1-w_R)\frac{k_C}{k}+w_R(1-w_R)(\pi_{C|D}-\pi_D)\frac{k_C}{k}\delta
    +(1-w_R)^2 (\pi_{C|D}-\pi_{D|D})\frac{k_C (k-k_C)}{k^2}\delta+\mathcal{O}(\delta^2),
\end{align}
where we conduct a second-order Taylor expansion at $\delta = 0$. We can utilize the first- or second-order terms later based on our requirements.

Summing over all $k_C$, we derive the probability that the number of $C$-players in the system increases by $1$ as follows:
\begin{align}\label{eq_pc+}
    \mathcal{P}\left(\Delta p_C=\frac{1}{N}\right)=&~(1-p_C)
    \sum_{k_C=0}^k{\frac{k!}{k_C!(k-k_C)!}q_{C|D}^{k_C} q_{D|D}^{k-k_C}\mathcal{P}(D\gets C)} \nonumber\\
    =&~(1-p_C)(1-w_R)q_{C|D}+(1-p_C)w_R(1-w_R)\left(
    \pi_{C|D}-(1-w_I)\frac{1+(k-1)q_{C|D}}{k}b
    \right)q_{C|D}\delta \nonumber\\
    &+(1-p_C)(1-w_R)^2 (\pi_{C|D}-\pi_{D|D})\frac{k-1}{k}q_{C|D}q_{D|D}\delta +\mathcal{O}(\delta^2),
\end{align}
which happens when a focal $D$-player is chosen with probability $1-p_C$ and adopts the strategy of a neighboring $C$-player in all possible scenarios involving $k_C$ cooperative neighbors. The summation is calculated using Eqs.~(\ref{bino2}), (\ref{bino4}), and conducting a second-order Taylor expansion, as the $\delta^0$ term will be eliminated later.

Upon the occurrence of the learning event $\mathcal{P}(D\gets C)$, the proportion of $CC$-edges in the system alters accordingly. Given $k_C$ cooperative neighbors surrounding the focal $D$-player, $k_C$ edges of $CD$ transform to $CC$-edges, and the proportion of $CC$-edges increases by $2k_C/(kN)$,
\begin{equation}
    \mathcal{P}\left(\Delta p_{CC}=\frac{2k_C}{kN}\right)=(1-p_C)\frac{k!}{k_C!(k-k_C)!} q_{C|D}^{k_C}q_{D|D}^{k-k_C} \mathcal{P}(D\gets C)\,.
\end{equation}
Summing over all possible values of $k_C$ yields the expected changes in the proportion of $CC$-edges:
\begin{align}\label{eq_pcc+}
    \sum_{k_C=0}^{k}{\frac{2k_C}{kN}\mathcal{P}\left(\Delta p_{CC}=\frac{2k_C}{kN}\right)}=&~
    (1-p_C)\sum_{k_C=0}^k{\frac{2k_C}{kN}\frac{k!}{k_C!(k-k_C)!}q_{C|D}^{k_C} q_{D|D}^{k-k_C}\left((1-w_R)\frac{k_C}{k}+\mathcal{O}(\delta)\right)} \nonumber\\
    =&~\frac{2(1-p_C)}{kN}(1-w_R)q_{C|D}[1+(k-1)q_{C|D}]+\mathcal{O}(\delta).
\end{align}
In this case, the summation is computed using Eq.~(\ref{bino3}), and we perform only a first-order Taylor expansion because, as we will observe later that the $\delta^0$ term will not be eliminated.

\subsection{Updating a $C$-player}
Next, we analyze the scenario where the focal agent is a $C$-player. Similarly, let us assume that there are $k_C$ cooperators surrounding this $C$-player. In this situation, the payoff for the $C$-player is given by:
\begin{equation}\label{eq_pic}
    \pi_C=w_I(-c+b)+\frac{1-w_I}{k}(-kc+k_C b)=-c+w_I b+(1-w_I)\frac{k_C}{k}b.
\end{equation}

The expected payoff for a $C$-player neighboring the focal $C$-player is
\begin{align}\label{eq_picc}
    \pi_{C|C}=&~w_I (-c+b)+\frac{1-w_I}{k}
    \left\{
    -c+b+\sum_{k_C'=0}^{k-1}{\frac{(k-1)!}{k_C'!(k-k_C'-1)!}q_{C|C}^{k_C'} q_{D|C}^{k-k_C'-1}[-(k-1)c+k_C' b]}
    \right\} \nonumber\\
    =&-c+w_I b+\frac{1-w_I}{k}[1+(k-1)q_{C|C}]b.
\end{align}
In this case, the focal $C$-player, as well as the remaining $k-1$ neighbors, leads to a cost $c$, while the focal $C$-player and $k_C'$ cooperators among the $k-1$ remaining neighbors bring a benefit $b$. The $C$-player also engages in the game with itself, with weight $w_I$.

The expected payoff for a $D$-player in the vicinity of the focal $C$-player is
\begin{align}\label{eq_pidc}
    \pi_{D|C}=&~\frac{1-w_I}{k}\left(b+
    \sum_{k_C'=0}^{k-1}{\frac{(k-1)!}{k_C'!(k-k_C'-1)!}q_{C|D}^{k_C'} q_{D|D}^{k-k_C'-1}k_C' b}\right) \nonumber\\
    =&~\frac{1-w_I}{k}[1+(k-1)q_{C|D}]b.
\end{align}

Analogous to Eq.~(\ref{eq_DgetsC}), the focal $C$-player transforms into a $D$-player with probability:
\begin{align}
    \mathcal{P}(C\gets D)=&~\frac{(1-w_R)/k\cdot (k-k_C) F_{D|C}}
    {w_R F_C+(1-w_R)/k\cdot [k_C F_{C|C}+(k-k_C)F_{D|C}]} \nonumber\\
    =&~(1-w_R)\frac{k-k_C}{k}+w_R(1-w_R)(\pi_{D|C}-\pi_C)\frac{k-k_C}{k}\delta \nonumber\\
    &+(1-w_R)^2 (\pi_{D|C}-\pi_{C|C})\frac{k_C (k-k_C)}{k^2}\delta+\mathcal{O}(\delta^2),
\end{align}
and the probability of the number of $C$-players in the system decreasing 1 is
\begin{align}\label{eq_pc-}
    \mathcal{P}\left(\Delta p_C=-\frac{1}{N}\right)=&~p_C
    \sum_{k_C=0}^k{\frac{k!}{k_C!(k-k_C)!}q_{C|C}^{k_C} q_{D|C}^{k-k_C}\mathcal{P}(C\gets D)} \nonumber\\
    =&~p_C(1-w_R)q_{D|C}+p_Cw_R(1-w_R)\left[
    \pi_{D|C}-\left(-c+w_Ib+(1-w_I)\frac{k-1}{k}q_{C|C}b\right)
    \right]q_{D|C}\delta \nonumber\\
    &+p_C(1-w_R)^2 (\pi_{D|C}-\pi_{C|C})\frac{k-1}{k}q_{C|C}q_{D|C}\delta +\mathcal{O}(\delta^2),
\end{align}
where a second-order Taylor expansion is performed for the same reason as in Eq.~(\ref{eq_pc+}).

If the learning event $\mathcal{P}(C\gets D)$ occurs, the proportion of $CC$-edges in the system decreases by $2k_C/(kN)$ when $k_C$ cooperative neighbors surround the focal $C$-player,
\begin{equation}
    \mathcal{P}\left(\Delta p_{CC}=-\frac{2k_C}{kN}\right)=p_C\frac{k!}{k_C!(k-k_C)!} q_{C|C}^{k_C}q_{D|C}^{k-k_C} \mathcal{P}(C\gets D).
\end{equation}

Taking into account all possibilities, the expected decrease in the proportion of $CC$-edges is given by
\begin{align}\label{eq_pcc-}
    \sum_{k_C=0}^{k}{\left(-\frac{2k_C}{kN}\right)\mathcal{P}\left(\Delta p_{CC}=-\frac{2k_C}{kN}\right)}=&
    -p_C\sum_{k_C=0}^k{\frac{2k_C}{kN}\frac{k!}{k_C!(k-k_C)!}q_{C|C}^{k_C} q_{D|C}^{k-k_C}\left((1-w_R)\frac{k-k_C}{k}+\mathcal{O}(\delta)\right)} \nonumber\\
    =&-\frac{2p_C}{kN}(1-w_R)(k-1)q_{C|C}q_{D|C}+\mathcal{O}(\delta).
\end{align}

\subsection{Diffusion approximation}
We can now formulate the system dynamics of $p_C$ and $p_{C|C}$ by employing the previously derived results. Utilizing Eqs.~(\ref{eq_pc+}) and (\ref{eq_pc-}), along with $p_{CD}=p_C q_{D|C}=(1-p_C) q_{C|D}$, we compute the instantaneous change in $p_C$ as
\begin{align}\label{eq_dpc}
    \dot{p}_C=&~
    \frac{1}{N}\mathcal{P}\left(\Delta p_C=\frac{1}{N}\right)+\left(-\frac{1}{N}\right)\mathcal{P}\left(\Delta p_C=-\frac{1}{N}\right) \nonumber\\
    =&~\frac{p_{CD}}{N}w_R(1-w_R)\left[(\pi_{C|D}-\pi_{D|C})+\left(-c+w_Ib
    +(1-w_I)\frac{-1+(k-1)(q_{C|C}-q_{C|D})}{k}b\right)\right]\delta \nonumber\\
    &+\frac{p_{CD}}{N}(1-w_R)^2 [(\pi_{C|D}-\pi_{D|D})q_{D|D}+(\pi_{C|C}-\pi_{D|C})q_{C|C}]\frac{k-1}{k}\delta +\mathcal{O}(\delta^2).
\end{align}
Observe that the $\delta^0$ terms are eliminated, and only non-zero terms remain from the $\delta^1$ terms. This is why we executed the Taylor expansion to the second-order $\delta^1$ in Eqs.~(\ref{eq_pc+}) and (\ref{eq_pc-}).

Analogously, using Eqs.~(\ref{eq_pcc+}) and (\ref{eq_pcc-}), we determine the instantaneous change in $p_{CC}$ as
\begin{align}\label{eq_dpcc}
    \dot{p}_{CC}=&\sum_{k_C=0}^{k}{\frac{2k_C}{kN}\mathcal{P}\left(\Delta p_{CC}=\frac{2k_C}{kN}\right)}+
    \sum_{k_C=0}^{k}{\left(-\frac{2k_C}{kN}\right)\mathcal{P}\left(\Delta p_{CC}=-\frac{2k_C}{kN}\right)} \nonumber\\
    =&~\frac{2p_{CD}}{kN}(1-w_R)[1+(k-1)(q_{C|D}-q_{C|C})]+\mathcal{O}(\delta).
\end{align}
Here, the $\delta^0$ term is non-zero, which is why we only need to perform the Taylor expansion to the first-order $\delta^0$ in Eqs.~(\ref{eq_pcc+}) and (\ref{eq_pcc-}).

Based on Eq.~(\ref{eq_dpcc}) and $q_{C|C}=p_{CC}/p_C$, we calculate the instantaneous change in $q_{C|C}$:
\begin{align}\label{eq_dqcc}
    \dot{q}_{C|C}=&~\frac{\mathrm{d}}{\mathrm{d}t}\left(\frac{p_{CC}}{p_C}\right) \nonumber\\
    =&~\frac{\dot{p}_{CC}p_C-\dot{p}_C p_{CC}}{{p_C}^2} \nonumber\\
    =&~\frac{2}{kN}\frac{p_{CD}}{p_C}(1-w_R)[1+(k-1)(q_{C|D}-q_{C|C})]+\mathcal{O}(\delta),
\end{align}
where we only employed the $\delta^0$ term in both $p_{CC}$ and $p_C$, as it is non-zero.

Comparing the governing equations, Eq.~(\ref{eq_dpc}) and Eq.~(\ref{eq_dqcc}), we observe that the change in $q_{C|C}$ illustrated by Eq.~(\ref{eq_dqcc}) is substantially faster than the change in $p_C$ portrayed by Eq.~(\ref{eq_dpc}). This is because the magnitude of $\dot{q}_{C|C}$ is $\delta^0$, while the magnitude of $\dot{p}_C$ is only $\delta^1$ in the $\delta\to 0^+$ limit, leading to the emergence of different time scales.

Owing to the distinct time scales, $q_{C|C}$ relaxes much faster than $p_C$. In other words, we can first determine the equilibrium of $q_{C|C}$ and then examine the dynamics of $p_C$ based on this foundation.

To compute the equilibrium of $q_{C|C}$, we solve $\dot{q}_{C|C}=0$ according to Eq.~(\ref{eq_dqcc}). The solution is $q_{C|C}-q_{C|D}=1/(k-1)$. Combining Eq.~(\ref{eq_otherv}), we can represent the remaining variables of the system by only $p_C$:
\begin{align}\label{eq_othervbypc}
    q_{C|C}&=\frac{k-2}{k-1}p_C+\frac{1}{k-1}, \nonumber\\
    q_{D|C}&=\frac{k-2}{k-1}(1-p_C), \nonumber\\
    p_{CD}&=\frac{k-2}{k-1}p_C(1-p_C), \nonumber\\
    q_{C|D}&=\frac{k-2}{k-1}p_C, \nonumber\\
    q_{D|D}&=1-\frac{k-2}{k-1}p_C.
\end{align}

According to Eq.~(\ref{eq_dpc}), we still need to calculate $\pi_{C|D}-\pi_{D|C}$, $(\pi_{C|D}-\pi_{D|D})q_{D|D}$, and $(\pi_{C|C}-\pi_{D|C})q_{C|C}$. Applying Eq.~(\ref{eq_othervbypc}) and considering Eqs.~(\ref{eq_picd}), (\ref{eq_pidd}), (\ref{eq_picc}), and (\ref{eq_pidc}), we obtain:
\begin{subequations}\label{eq_piminusbypc}
\begin{align}
    \pi_{C|D}-\pi_{D|C}=&-c+w_I b+\frac{1-w_I}{k}(k-1)q_{C|C}b-\frac{1-w_I}{k}[1+(k-1)q_{C|D}]b \nonumber\\
    =&-c+w_I b,
\end{align}
\begin{align}
    (\pi_{C|D}-\pi_{D|D})q_{D|D}=&\left[-c+w_I b+\frac{1-w_I}{k}(k-1)q_{C|C}b-\frac{1-w_I}{k}(k-1)q_{C|D}b\right]
    \left(1-\frac{k-2}{k-1}p_C\right) \nonumber\\
    =&\left(-c+w_I b+\frac{1-w_I}{k}b\right)\left(1-\frac{k-2}{k-1}p_C\right),
\end{align}
\begin{align}
    (\pi_{C|C}-\pi_{D|C})q_{C|C}=&\left\{-c+w_I b+\frac{1-w_I}{k}[1+(k-1)q_{C|C}]b-\frac{1-w_I}{k}[1+(k-1)q_{C|D}]b\right\} \nonumber\\
    &\times\left(\frac{k-2}{k-1}p_C+\frac{1}{k-1}\right) \nonumber\\
    =&\left(-c+w_I b+\frac{1-w_I}{k}b\right)\left(\frac{k-2}{k-1}p_C+\frac{1}{k-1}\right).
\end{align}
\end{subequations}

Substituting Eqs.~(\ref{eq_othervbypc}) and (\ref{eq_piminusbypc}) into Eq.(\ref{eq_dpc}), we can express $\dot{p}_C$ solely in terms of $p_C$:
\begin{equation}
    \dot{p}_C=\frac{k-2}{(k-1)N}p_C(1-p_C)(1-w_R)\left\{
    -(1+w_R)c+[(k-1)w_I+(k+1)w_I w_R+1-w_R]b/k
    \right\}\delta +\mathcal{O}(\delta^2),
\end{equation}
which bears a resemblance to the replicator dynamics in well-mixed populations. There are two equilibria, ${p_C}^*=0$ and ${p_C}^{**}=1$. In accordance with the standard stability analysis~\cite{taylor1978evolutionary}, when $b/c<(b/c)^\star$, the system is stable at ${p_C}^*$, and defection prevails. When $b/c>(b/c)^\star$, the system is stable at ${p_C}^{**}$, and cooperation prevails. Here,
\begin{equation}
    \left(\frac{b}{c}\right)^\star=\frac{1+w_R}{(k-1)w_I+(k+1)w_I w_R+1-w_R}k, 
\end{equation}
which is consistent with the results of the IBD method when $N\to +\infty$. Finally, to rigorously complete the theoretical deduction, we proceed to calculate the fixation probability.

\subsection{Fixation probability}
To determine the fixation probability, one must solve the Kolmogorov backward equation~\cite{karlin1981second,matsuda1992statistical,ewens2004mathematical}. We begin by defining the following two quantities:
\begin{align}
    \text{E}(\Delta p_C)\simeq&\left[\frac{1}{N}\mathcal{P}\left(\Delta p_C=\frac{1}{N}\right)+\left(-\frac{1}{N}\right)\mathcal{P}\left(\Delta p_C=-\frac{1}{N}\right)\right]\Delta t \nonumber\\
    =&~\frac{k-2}{(k-1)N}p_C(1-p_C)(1-w_R)\left\{
    -(1+w_R)c+[(k-1)w_I+(k+1)w_I w_R+1-w_R]b/k
    \right\}\delta\Delta t \nonumber\\
    \equiv&~m(p_C)\Delta t,
\end{align}
\begin{align}
    \text{Var}(\Delta p_C)\simeq&\left[\left(\frac{1}{N}\right)^2\mathcal{P}\left(\Delta p_C=\frac{1}{N}\right)+\left(-\frac{1}{N}\right)^2\mathcal{P}\left(\Delta p_C=-\frac{1}{N}\right)\right]\Delta t \nonumber\\
    =&~\frac{2(k-2)}{(k-1)N^2}p_C(1-p_C)(1-w_R)\Delta t \nonumber\\
    \equiv&~v(p_C)\Delta t.
\end{align}
Subsequently, we obtain
\begin{equation}
    -\frac{2m(p_C)}{v(p_C)}=-\frac{N}{k}
    \left\{
    -(1+w_R)kc+[(k-1)w_I+(k+1)w_I w_R+1-w_R]b
    \right\}\delta,
\end{equation}
and 
\begin{align}
    G(p_C)=&~\exp{\left(
    -\int{\frac{2m(p_C)}{v(p_C)}}\,\mathrm{d}p_C
    \right)} \nonumber\\
    =&~\exp{\left(
    -\frac{N}{k}
    \left\{
    -(1+w_R)kc+[(k-1)w_I+(k+1)w_I w_R+1-w_R]b
    \right\}\delta p_C+C_0
    \right)} \nonumber\\
    =&\left(
    1-\frac{N}{k}
    \left\{
    -(1+w_R)kc+[(k-1)w_I+(k+1)w_I w_R+1-w_R]b
    \right\}\delta p_C
    \right)\tilde{C}_0 +\mathcal{O}(\delta^2),
\end{align}
where $C_0$ and $\tilde{C}_0=\exp{C_0}$ are constants arising from integral calculations.

We denote the initial cooperation level at $t_0$ as $p_C(t_0)$, and the fixation probability of cooperation, starting with the cooperation level $p_C(t_0)$, as $\phi_C[p_C(t_0)]$. To determine $\phi_C[p_C(t_0)]$, we solve the following equation:
\begin{equation}
    0=m[p_C(t_0)]\frac{\mathrm{d}\phi_C [p_C(t_0)]}{\mathrm{d}p_C(t_0)}+
    \frac{v[p_C(t_0)]}{2}
    \frac{\mathrm{d}^2\phi_C [p_C(t_0)]}{\mathrm{d}{p_C}^2(t_0)}
\end{equation}
with boundary conditions $\phi_C(0)=0$, $\phi_C(1)=1$. The solution is
\begin{align}\label{eq_phic}
    \phi_C[p_C(t_0)]=&~\frac{\int_{0}^{p_C(t_0)}{G(p_C)}\,\mathrm{d}p_C}{\int_{0}^{1}{G(p_C)}\,\mathrm{d}p_C} \nonumber\\
    =&~\frac{\left.\left(
    p_C-\dfrac{N}{2k}
    \left\{
    -(1+w_R)kc+[(k-1)w_I+(k+1)w_I w_R+1-w_R]b
    \right\}\delta {p_C}^2
    \right)\tilde{C}_0 \right|_{0}^{p_C(t_0)}}
    {\left.\left(
    p_C-\dfrac{N}{2k}
    \left\{
    -(1+w_R)kc+[(k-1)w_I+(k+1)w_I w_R+1-w_R]b
    \right\}\delta {p_C}^2
    \right)\tilde{C}_0 \right|_{0}^{1}} \nonumber\\
    =&~p_C(t_0)+p_C(t_0)[1-p_C(t_0)]
    \frac{\dfrac{N}{2k}\left\{
    -(1+w_R)kc+[(k-1)w_I+(k+1)w_I w_R+1-w_R]b
    \right\}\delta}{1-\dfrac{N}{2k}\left\{
    -(1+w_R)kc+[(k-1)w_I+(k+1)w_I w_R+1-w_R]b
    \right\}\delta} \nonumber\\
    =&~p_C(t_0)+\frac{p_C(t_0)[1-p_C(t_0)]}{2}N\left\{
    -(1+w_R)c+[(k-1)w_I+(k+1)w_I w_R+1-w_R]b/k
    \right\}\delta +\mathcal{O}(\delta^2).
\end{align}
Since $N\to +\infty$, we can approximate $N_C/N\approx p_C(t_0)$ and $(N-N_C)/(N-1)\approx 1-p_C(t_0)$. In this manner, the fixation probability calculated by Eq.~(\ref{eq_phic}) is equivalent to the cooperation level given by Eq.~(\ref{eq_rhodg}), that is, $\phi_C=\rho_C$, in the limit of $N\to +\infty$.

As discussed in the introductory paragraph of Section~\ref{sec_theo}, cooperation dominates with probability $p_C(t_0)$ under neutral drift, and evolution favors cooperation under weak selection if the probability of cooperation dominance exceeds $p_C(t_0)$. This implies that $\phi_C[p_C(t_0)] > p_C(t_0)$. According to Eq.~(\ref{eq_phic}), $\phi_C[p_C(t_0)] > p_C(t_0)$ necessitates
\begin{equation}
    \frac{b}{c}>\frac{1+w_R}{(k-1)w_I+(k+1)w_I w_R+1-w_R}k, 
\end{equation}
which constitutes a generalization of the well-known $b/c > k$ rule~\cite{ohtsuki2006simple} and is consistent with the result obtained by the IBD method.

\twocolumn
% %% Loading bibliography style file
% % \bibliographystyle{model1-num-names}
% \bibliographystyle{elsarticle-num-names}
% \bibliography{cas-refs}

\end{document}